\documentclass[twocolumn]{aastex63}
\usepackage{amsmath}	
\usepackage{amssymb}	
\usepackage{rotating}

\definecolor{asparagus}{rgb}{0.0, 0.5, 0.0}

\received{March 1, 2023}
\revised{April 1, 2021}
\accepted{\today}
\submitjournal{ApJ}

\shorttitle{3D spectroscopy of the triple AGN}
\shortauthors{Ben\'itez et al.}

\usepackage{epstopdf}
\usepackage{ulem}
\usepackage{amssymb}
\usepackage{times}
\usepackage{graphicx}
\usepackage[usenames,dvipsnames]{pstricks}
\usepackage{psfrag}
\usepackage{lipsum}
\usepackage{natbib}
\usepackage{textcase}

\def\apjl{ApJ}
\def\apjs{ApJS}
\def\ao{Appl.~Opt.}

\newcommand{\be}{\begin{equation}}
\newcommand{\ee}{\end{equation}}
\newcommand{\bary}{\begin{eqnarray}}
\newcommand{\eary}{\end{eqnarray}}

\def\ergs{erg s$^{-1}$}

\def\ha{H$\alpha$}

\def\hb{H$\beta$}

\def\oi{{\sc [Oi]}}
\def\oi{{[O\sc{i}]}\/}

\def\oiii{{[O\sc{iii}]}\/}

\def\nii{{[N\sc{ii}]}\/}
\def\niia{{[N\sc{ii}]}$\lambda$6548\/}
\def\niib{{[N\sc{ii}]}$\lambda$6583\/}

\def\sii{{[S\sc{ii}]}$\lambda\lambda$6716,6731\/}

\def\d{$\delta$}
\def\l{$\lambda$}
\def\s{$\sigma$}
\def\t{$\theta$}
\def\kms{km\,s$^{-1}$\,}
\def\D{$\Delta$}

\def\mbh{M$\rm_{BH}$}

\def\l5100{L$_{\it 5100}$}
\def\J1027{J102700.40+174900.8}
\def\objedos{J102700.55+174900.2}
\def\objetres{J102700.38+174902.6}
\def\meg{{\it MEGARA}}

\begin{document}

\title{3D spectroscopy with GTC-MEGARA of the triple AGN candidate in SDSS J102700.40+174900.8}

\author[0000-0003-1018-2613]{Erika Ben\'itez}
\affiliation{Universidad Nacional Aut\'onoma de M\'exico, Instituto de Astronom\'ia, AP 70-264, CDMX 04510, Mexico}

\author[0000-0002-9790-6313]{H\'ector Ibarra-Medel}
\affiliation{Escuela Superior de F\'{\i}sica y Matem\'aticas, Instituto Polit\'ecnico Nacional, U.P. Adolfo L\'opez Mateos, C.P. 07738, Ciudad de M\'exico, M\'exico}
\affiliation{Universidad Nacional Aut\'onoma de M\'exico, Instituto de Astronom\'ia, AP 70-264, CDMX 04510, Mexico}

\author[0000-0002-1656-827X]{Castalia Alenka Negrete}
\affiliation{Conacyt Research Fellow at Universidad Nacional Aut\'onoma de M\'exico, Instituto de Astronom\'ia, AP 70-264, CDMX 04510, Mexico}

\author[0000-0002-2653-1120]{Irene Cruz-Gonz\'alez}
\affiliation{Universidad Nacional Aut\'onoma de M\'exico, Instituto de Astronom\'ia, AP 70-264, CDMX 04510, Mexico}

\author[0000-0002-0674-1470]{Jos\'e Miguel Rodr\'iguez-Espinosa}
\affiliation{Instituto de Astrof\'isica de Canarias, V\'ia L\'actea, s/n, 38205, La Laguna, Spain}
\affiliation{Departamento de Astrof\'isica, Universidad de La Laguna (ULL), 38205, Spain}

\author[0000-0003-0049-5210]{Xin Liu}
\affiliation{Department of Astronomy, University of Illinois at Urbana-Champaign, Urbana, IL 61801, USA}
\affiliation{National Center for Supercomputing Applications, University of Illinois at Urbana-Champaign, Urbana, IL 61801, USA}

\author[0000-0003-1659-7035]{Yue Shen}
\affiliation{Department of Astronomy, University of Illinois at Urbana-Champaign, Urbana, IL 61801, USA}
\affiliation{National Center for Supercomputing Applications, University of Illinois at Urbana-Champaign, Urbana, IL 61801, USA}

\correspondingauthor{E. Ben\'itez.}
\email{erika@astro.unam.mx}

\begin{abstract}
Triple AGN systems are expected to be the result of the hierarchical model of galaxy formation. Since there are very few of them confirmed as such, we present the results of a new study of the triple-AGN candidate SDSS \J1027 (center nucleus) through observations with \textit{GTC}-\meg\ Integral Field Unit. 1D and 2D analysis of the line ratios of the three nuclei allow us to locate them in the EW(\ha) vs. \nii /\ha\ diagram. The central nucleus is found to be a retired galaxy (or fake AGN). The neighbors are found to be a strong AGN (southeastern nucleus, \objedos) compatible with a Sy2 galaxy, and a weak AGN (northern nucleus, \objetres) compatible with a LINER2. We find evidence that the neighbors constitute a dual AGN system (Sy2-LINER2) with a projected separation of 3.98\,kpc in the optical bands. The \ha\ velocity map shows that the northern nucleus has an \ha\ emission with a velocity offset of $\sim$-500\,\kms, whereas the southeastern nucleus has a rotating disk and \ha\ extended emission at kpc scales. Chandra archival data confirm that the neighbors have X-ray (0.5-2)\,keV and (2-7)\,keV emission, whereas the center nucleus shows no X-ray emission. A collisional ring with knots is observed in the HST images of the southeastern nucleus. These knots coincide with star formation regions that along with the ring are predicted in a head-on collision. In this case, the morphology changes are probably due to a minor merger that was produced by the passing of the northern through the southeastern nucleus. 
\end{abstract}



\section{Introduction} 
\label{section:intro} 

Galaxy mergers are major events that play an important role in galaxy evolution within the hierarchical Lambda cold dark matter \citep[$\Lambda$CDM,][]{1978MNRAS.183..341W,1991ApJ...379...52W,2005ApJ...620L..79S} cosmology. They are found to be involved in the triggering of bursts of star formation
\citep[e.g.][]{1991ApJ...370L..65B,1996ApJ...471..115B,2004MNRAS.350..798B,2010MNRAS.407...43C,2013MNRAS.430.1901H,2018MNRAS.479.3952B} and in the accretion of material onto a central supermassive black hole 
\citep[SMBH,][]{2005Natur.433..604D,2008ApJ...676...33D,2006ApJS..163....1H,2008ApJS..175..356H}.

Since it is well established that almost all massive galaxies harbor a central supermassive black hole \citep[SMBH,][]{1995ARA&A..33..581K,1998AJ....115.2285M,2000ApJ...539L...9F, 2000ApJ...539L..13G}, then 
galaxy interactions can get closer two or more SMBHs before the final coalescence phase.

Dual Active Galactic Nuclei (DAGN) are thus expected to be found in the framework of galaxy hierarchical evolutionary models. DAGN harbor two active SMBHs and simulations show that are systems in an intermediate evolutionary stage between the first encounter and final coalescence of two merging gas-rich galaxies \citep[see][for a review]{2019NewAR..8601525D}. In this evolutionary phase, the two SMBHs show a spatial projected separation of 1-100\,kpc.\ This phase can last $\sim$\,100 million years \citep[see][]{2017MNRAS.469.4437C} and during this time, single or both SMBHs show AGN activity \citep{2012ApJ...748L...7V,2013MNRAS.429.2594B,2018MNRAS.478.3056B}.

DAGN are hard to detect, mainly due to their small (sometimes, $<$ 1\arcsec) spatial projected separations. So, multiwavelength studies, particularly observations in the radio and X-ray bands, result ideal to confirm them as such \citep[see, e.g.][]{2019MNRAS.490.5521B,2022MNRAS.516.5270B}. In the optical bands, DAGN candidates can be selected using data from surveys like the Sloan Digital Sky Survey (SDSS). Images showing two or more nuclei and with double-peaked narrow emission lines in their spectra 
\citep{2009ApJ...705L..76W,2012ApJS..201...31G,2010ApJ...716..866S,Liu+2010,2020ApJ...904...23K} can be related to a pair of SMBHs that have kpc-scale separations with the emission from their NLR overlapped. We will refer to double-peaked narrow emission line AGN as DPAGN.

Triple AGN can also form during the merger process if, for example, DAGN merges with another AGN in their way, and the three are simultaneously accreting. However, there are very few triple AGN systems confirmed up to now \citep[][]{2019ApJ...883..167P,2021A&A...651L...9Y}. Determining the rate of dual AGN or multiple AGN in galaxy mergers will explore merger-induced activity in AGN and SMBH growth and evolution.

SDSS \J1027 (hereafter, we will refer to it as  
\J1027) appears classified as a galaxy 
in the SDSS-DR7 \citep{2009ApJS..182..543A} with a redshift z=0.066\footnote{At z=0.066, the spatial scale of 1\arcsec\, 
corresponds to 1.274 kpc \citep{2006PASP..118.1711W}.}. \J1027\, visually shows in the optical bands that it is part of three intense and apparently nearby nuclear sources. In a previous work done by \citet{2011ApJ...736L...7L}, hereafter L11, they show that it is a system of three emission-line nuclei, two of which are offset by 450 and 110\,\kms in velocity and by 2.4 and 3.0 kpc in projected separation from the central nucleus (i.e., from \J1027). Therefore, the authors present evidence of a kpc scale triple AGN system. Also, they have found that the three nuclei are obscured AGN based on optical diagnostic emission line ratios and obtained estimates of their black hole masses using the M$_{BH}$ -$\sigma_{\star}$ correlation given in \citet{2002ApJ...574..740T}. Based on the BPT diagnostic diagrams \citep[see][]{1981PASP...93....5B,1987ApJS...63..295V}, L11 find that source \J1027\ (identified as A in their Fig.1) is in the locus of Low-Ionization Narrow Emission Region, i.e., it is a LINER source \citep{1980A&A....87..152H}. In the case of \objedos\ (source B, at the SE of A) it is found in the locus between AGN-HII region, and source \objetres\ (source C, at the N of A) appears in the locus between Seyfert-LINER galaxies. 

In this work, we present an optical and kinematical study done on the triple AGN candidate \J1027\, through  observations carried out with the Gran Telescopio Canarias (GTC) and the Multi-Espectr\'ografo de Alta Resoluci\'on para Astronom\'ia (\meg) instrument. With \meg, we apply the Integral Field Spectroscopy (IFS) observational technique, which can spatially resolve the spectral information of an object with the instrumental implementation of the integral field units \citep[IFU, see][for a review]{Sanchez+2020}. Therefore, with IFS data, it is possible to spectroscopically disentangle the associated sources to \J1027. We also present the analysis of high-resolution $U$- and $Y$-band imaging from the {\it Hubble Space Telescope}\footnote{All the {\it HST}, data used in this paper can be found in MAST: \dataset[10.17909/bkz9-zr94]{http://dx.doi.org/10.17909/bkz9-zr94}.} (HST), data from the Legacy SDSS survey \citep{York+2000}, and archival data of the {\it Chandra} X-ray space telescope. In section~\ref{section:obs} observations and data processing are presented. Data analysis appears in section~\ref{section:ana}. Results and discussion are given in section~\ref{section:res} and  section~\ref{section:dis}. Finally, conclusions are presented in section~\ref{section:conc}. The cosmology adopted in this work is
$H_{0}$\,=\,69.6~km/,s$^{-1}$\,Mpc$^{-1}$, $\Omega_{m}$\,=\,0.286 and $\Omega_{\lambda}$\,=\,0.714 \citep{2014ApJ...794..135B}.

\section{Observations and data processing}
\label{section:obs}

\subsection{MEGARA}
\label{subsection:megara}
The instrument MEGARA is located in the Cassegrain focus of the GTC \citep[see][]{2018SPIE10702E..16C}. We used the IFU mode of MEGARA, also called \textit{Large Compact Bundle} (LCB), that provides a field of view of 12.5\arcsec $\times$ 11.3\arcsec. Observations were done in GTC3-21AMEX and GTC8-21BMEX (PI: Ben\'itez) in order to cover the blue and red spectral ranges. The three nuclei in \J1027, see left panel of Figure~\ref{fig:regions} and Table~\ref{tab:identify}, were observed. In all blocks, the position of the telescope was centered at \J1027. This position guaranteed that the three nuclei were inside the Field of View (FoV) of \meg. Details on the  observations are shown in Table~\ref{tab:log}.

\subsection{Processing the data cubes}
\label{subsection:datacubes}

\subsubsection{2D Reduction}

We use the \meg\, official Data Reduction Pipeline (MEGARA DRP\footnote{\url{https://megaradrp.readthedocs.io/en/stable/}}) to perform the  data reduction. The MEGARA DRP performs the standard reduction of the raw multi-fiber exposures: bias subtraction, fiber extraction, fiber flat-field correction, flexure correction, 
wavelength calibration, and flux calibration. For the flux calibration, we use the spectrophotometry standard stars HR5501 and HR1544 from \citet{Oke+1990} and 
\citet{Hamuy+199} catalogs. The final output for the MEGARA DRP is a fits file that 
contains the science-calibrated row-stack-spectra (RSS) for each fiber, with metadata 
that has the relative fiber positions to the IFU center. In addition, we correct our wavelength solution to have a radial velocity in the heliocentric reference frame.

\subsubsection{3D Reduction}

With the flux and wavelength calibrated RSS, we proceed to re-arrange and reconstruct the final data cube with a homogeneous spatially grid sampling. To create the spatially re-sampling data cube, we apply the methodology described in \citet{Ibarra-Medel+2019}, which is based on the same methodology used in MaNGA \citep{Law:2016aa} and CALIFA \citep{sanchez12a} surveys. The method interpolates the flux at each spaxel of a given spatial grid using the next expression:  
\begin{equation}
F_{x,y}(\lambda)=\sum_i^{n_{fib}}w(\lambda,r_i)F_{i}(\lambda)/\sum_i^{n_{fib}}w(\lambda,r_i),
\label{spatial_sampling}
\end{equation}
where the value of $F_{i}(\lambda)$ is the 1d-spectra from the $i-th$ fiber from the calibrated RSS. The weight $w(\lambda,r)$ is a Gaussian kernel defined as $w(\lambda,r)=exp(-0.5(r_{\lambda}/\sigma)^2)$, with $r$ as the distance from the spaxel ($x,y$) to the center of the fiber $i$ at wavelength $\lambda$. We set a value of $\sigma=0.31$\arcsec, to obtain a weighting kernel with a full-width half maximum (FWHM) comparable with the diameter size of the $\meg$ fiber size (0.62\arcsec). Finally, we define a final spaxel scale to be $0.35$\,\arcsec/spaxel to avoid any spatial oversampling once considering the fiber diameter size. 

\subsubsection{Atmospheric differential refraction}

We consider the wavelength dependency of the atmospheric differential refraction (ADR) on the cube reconstruction. To implement a correction of the ADR, we first use the international atomic time (TAI) of each exposure to estimate the effective TAI for the total OB. Then, we use the exact position of the GTC observatory to estimate the effective zenith angle ($z$) of the OB. Hence, we calculate the ADR shift with the use of the \cite{Filippenko:1982aa} formulation, based on the \citet{Smart:1931aa} formula:
\begin{multline}
    \Delta R(\lambda,\lambda_0,z,T,P,f)=
    \\180\times3600\frac{R(\lambda,T,P,f)-R(\lambda_0,T,P,f)}{10^6\pi}\tan z, \\
    R(\lambda,T,P,f)=(n(\lambda)_{T,P,f}-1)\times10^6.\label{ADR_e}
\end{multline}
The value of $n$ is the refraction index of the air as a function of the wavelength. The parameters $T$,$P$  and $f$ indicate the dependency of the air temperature, atmospheric pressure, and water vapor pressure of the observational site \citep{Filippenko:1982aa}. $\Delta R(\lambda,\lambda_0,z,T,P,f)$ is the shift (in arcseconds) of the target at a specific wavelength in reference to the wavelength $\lambda_0$. The shift direction in reference to the north celestial pole is given by the parallactic angle. Therefore, we can accurately measure the spatial shift per wavelength in reference to the center position of the IFU, and correct this shift in Equation \ref{spatial_sampling} for all wavelength sampling. The air temperature, atmospheric pressure, and water vapor pressure are taken from the observational log information and from the headers of the fits raw exposures. 

\subsubsection{VPH co-adding}

At this step, we obtain four data cubes, one per VPH/OB. We proceed to co-add the four data cubes in one single data cube that has a spectral coverage from 4340 to 7303 \AA. To do so, we use the surface brightness distribution of the three nuclei to estimate their central positions per each OB within the data cubes. We estimate the shifts between the three centroids of the OB with the VPH675 and the centroids between the remaining OBs. With these shifts, we can calculate a 2D transformation to spatially shift and rotate the VPH570, VPH521, and VPH480 data cubes to be exactly at the same spatial position within the FoV of the data cube of the VPH675. In this step, we implement a 2D surface interpolation of the spatial grid per each spectral sampling. The interpolation generates a new set of data cubes for the VPH570, VPH521, and VPH480 with the same spatial grid and FoV of the VPH675 data cube. Then, we define a new data cube with a total spectral range of the combined VPH's (4340 to 7303 \AA). In this case, we set the spectral sampling to 0.25 \AA\ per pixel, and we degrade the spectra resolution for all the data cubes to R=5000 (VPH 480), which is the lowest resolution of the four VPHs. To degrade the spectral resolution, we convolved the spectra with a Gaussian filter that has an FWHM defined as $\Delta \lambda^2=\Delta \lambda_{5000}^2-\Delta \lambda_{0}^2$, with $\Delta \lambda_{5000}$ as the spectral resolution of the VPH 480, and $\Delta \lambda_{0}$ as the original spectral resolution of the degraded spectra. Finally, we interpolate the spectra of each data cube within the new spectral sampling. On the  spectral regions where two or more VPH overlap, we obtain the average flux among them. Therefore, we obtain a single data cube that merges the four OBs. 

\begin{figure*}[ht]
\centering
\includegraphics[height=5.5cm]{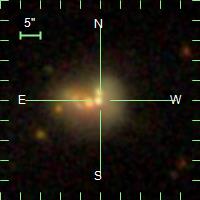}
\includegraphics[height=5.5cm]{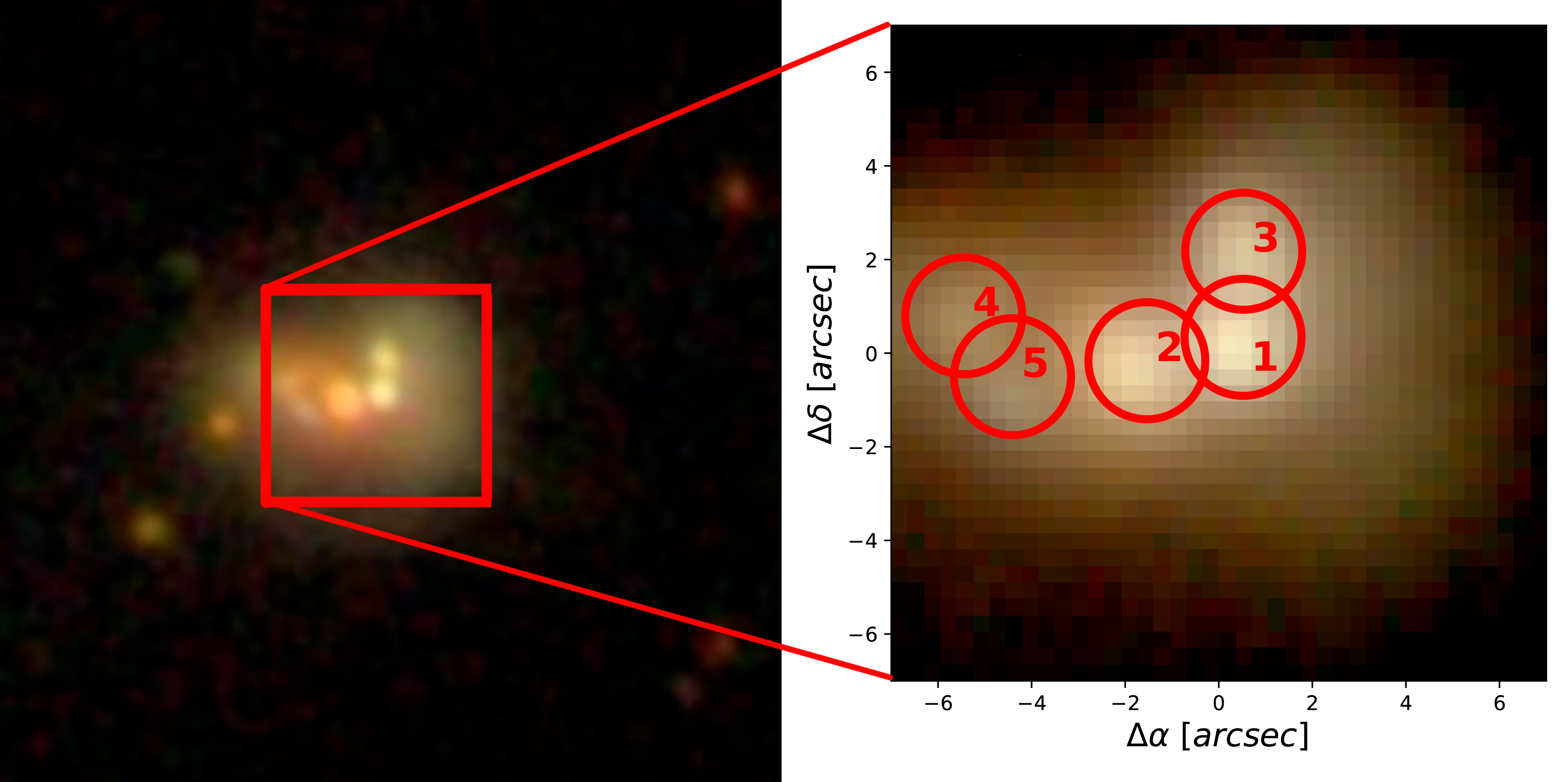}
\caption{Left panel: SDSS-DR7 image centered at \J1027\, obtained at MJD 53710 (2005 12 06). The image shows the three nuclear sources, one at the center coordinates of \J1027\ and two more, one at the North and the other at the SE. See Table~\ref{tab:identify} for target identifications. Central panel: The same as in the left image, but showing in a red box the \meg\ FoV. Right panel: The reconstructed rgb image from the \meg\ IFU cube. The red circles show the five selected regions that were analyzed in this work, all of them with an aperture of 2.5 arcsecs.}
\label{fig:regions}
\end{figure*}

\subsubsection{The Noise spectrum }

\begin{figure*}
\centering
\includegraphics[width=2.0\columnwidth]{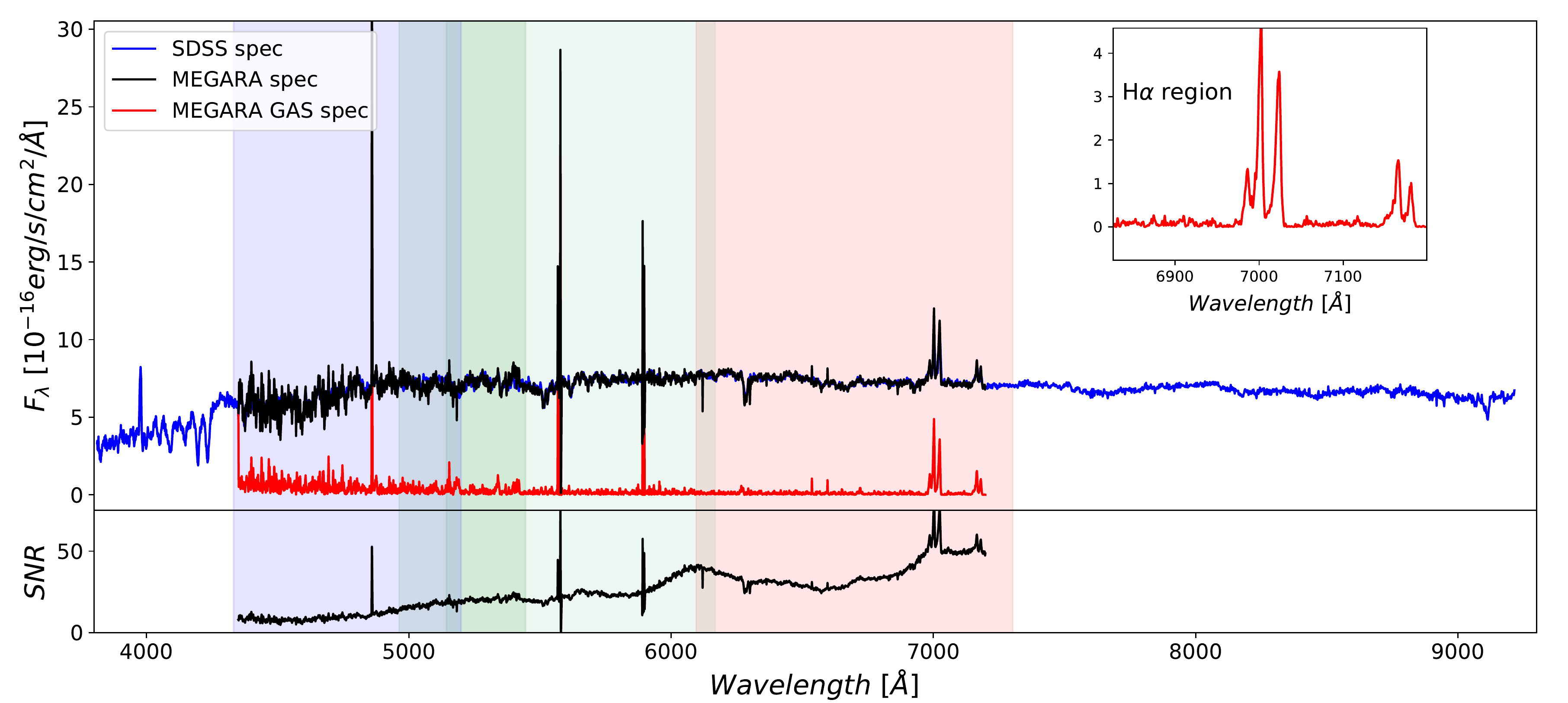}
\caption{In the upper panel, we show in black the \meg \ spectrum extracted with a fiber aperture of 3\arcsec, centered in the fiber position of the SDSS archival spectrum of \J1027\ obtained at MJD 54140 (2007 02 09). The archival SDSS spectrum appears overlapped in blue. In red, we show the gas spectra obtained after subtracting the contribution from the stellar population using {\sc pyPipe3D}. The shaded color regions represent the \meg\, VPH spectral regions: blue for the VPH480 LR-B, dark green for the VPH521 MR-G, light green for the VPH570 LR-V, and red for the VPH675 LR-R. The inset panel shows a zoom-in of the \ha\ region for the gas spectra. The lower panel shows the SNR of the \meg \ spectrum.}
\label{spec}
\end{figure*}

To estimate the noise spectra of the final cube, we smooth the resolved spectrum spaxel by spaxel with a Gaussian kernel of $\sigma=100$ \AA, in order to damp the contribution of the strong emission lines and to obtain a smooth spectrum that traces the continuum. With the smoothed spectrum, we obtain a residual spectrum defined as the original minus the smoothed spectrum. With the residual spectrum, we calculate the 1-$\sigma$ percentile values (25th to 75th percentile) of the spectral residuals within a shifting window of 800 \AA. This window is centered at each spectral sampling of the spectrum; therefore, we obtain the 1-$\sigma$ percentile of the residuals at each wavelength. The use of the 1-$\sigma$ percentile rejects the outliers associated with the emission lines and gives the effective variance associated with the observed spectra. In Figure \ref{spec}, we show an example of the final Signal to Noise ratio (SNR) spectra within our reduced data cube. We recuperate an average SNR of $\approx$\,60 on the VPH675 spectral region. This VPH has the largest exposure time for our observational campaign. The lowest SNR corresponds to the VPH480, with an average SNR of $\approx$\,7. Unfortunately, the VPH480 spectral region that was selected to observe the $H\beta$ complex is obscured by dust, and the SNR in this spectral region is not enough to achieve a good detection.

\subsubsection{Astrometry}

With the merged data cube, we now proceed to calculate accurate astrometry. We use the SDSS r-band photometric image of the region of J102700.40+174900.8. We calculate the centroids of the three nuclei within the SDSS photometric image and compare them with the centroids of the same three nuclei within the FoV of the data cube. With this information, we obtained the World Coordinate System (WCS) for the merged data cube to match the WCS of the r-band SDSS photometric image. The astrometric positions and redshifts of the selected five regions are shown in Table~\ref{tab:astrometry} and the projected separations between the three nuclei are given in Table~\ref{tab:projsep}.

\subsubsection{Second order spectrophotometric correction}

We use the SDSS $g$ and $r$ photometric bands to create a second-order correction on the spectrophotometry calibration. The SDSS $r$-band and most of the $g$-band fit within the merged data cube spectral range of 4340 to 7303 \AA. Therefore, we can calculate the synthetic $r$ and $g$ band photometric fluxes from the data cube. With the synthetic fluxes, we measure any flux deviation from the SDSS photometry and calculate a 2D flux correction for our data cube to reproduce the SDSS g-r colors. Finally, we correct the final data cube by the galactic extinction using the \citet{Schlafly+2011} galactic dust extinction maps with the \citet{Fitzpatrick+1999} extinction law.

\subsubsection{Point Spread Function characterization}

To characterize the effective Point Spread Function (PSF) within our final \meg \ data cube, we first apply the same reduction procedure (described above) to the standard star HR1544 to recuperate its data cube. The effective PSF on an IFU observation is the total contribution of the Telescope seeing and the 2D reconstruction of the data cube \citep{Ibarra-Medel+2019,Sanchez+2020}.  With the data cube of HR1544, we perform a surface brightness ($\mu$) profile fitting following the methodology described in Ibarra-Medel et al. in prep and in \citet{Ibarra-Medel+2022b}. The methodology first integrates the cube along the spectral axis to obtain a high-resolution $\mu$ map. Then, we perform the $\mu$ profile fitting with the use of a Moffat profile \citep{Moffat+1969,Trujillo+2001}. For completeness, we also explore a Gaussian $\mu$ profile. The $\mu$ fitting uses a Markov chain Monte Carlo (MCMC) method. We find that the effective PSF of our reconstructed data cube has a Moffat profile with an FWHM of 1.25\arcsec with a $\beta$ value of 9.32. The $\beta$ value indicates the extension of the PSF wings. When $\beta$ asymptotically tends to infinite, the Moffat profile tends to a Gaussian shape. If we fit a Gaussian profile directly, we obtain a PSF FWHM of 0.87\arcsec. The difference between the Gaussian fit and the Moffat fit is due to the PSF tail contribution to the profile. Therefore, we use the Moffat fitting to characterize the effective PSF of our data cube. This gives us an effective spatial resolution of 1.25\arcsec within the cube.

\begin{deluxetable}{ccc}
\tablenum{1}
\tablecaption{Target identification
\label{tab:identify}}
\tablewidth{0pt}
\tablehead{
\colhead{Target} & \colhead{L11} & \colhead{This work} 
}
\decimalcolnumbers
\startdata
SDSS J102700.40+174900.8 & A & 1 \\
SDSS J102700.55+174900.2 & B & 2 \\
SDSS J102700.38+174902.6 & C & 3 \\
\enddata
\tablecomments{The three principal nuclei identified with the nomenclature previously used by \citet{2011ApJ...736L...7L} and, for simplicity, in the images shown in this work.}
\end{deluxetable}

\begin{deluxetable*}{ccccccccc}
\tablenum{2}
\tablecaption{Log of observations}
\label{tab:log}
\tablewidth{0pt}
\tablehead{
\colhead{Date Obs} & \colhead{VPH} & \colhead{R} & \colhead{Spectral range} & \colhead{Exposure time} & \colhead{Moon} & \colhead{sky} &\colhead{airmass} & \colhead{seeing} 
\\
\colhead{} & \colhead{} & \colhead{} & \colhead{\AA} & \colhead{s} & \colhead{} & \colhead{} &\colhead{\arcsec}
}
\decimalcolnumbers
\startdata
2021 March 2 & 675LR-R & 5900 & 6096.54-7303.21 &
1860 & grey & clear & 1.2 & 0.6\\
2022 January 1 & 570LR-V & 5850 & 5143.74-6168.19 &
1116 & dark & clear & 1.5 & 1\\
2022 March 4 & 521MR-G & 12000 & 4963.22-5445 & 900 & dark & clear & 1.1 & 1.1\\
2022 March 5 & 480LR-B & 5000 & 4332.05-5199.96 & 900 & dark & clear & 1.1 & 1.1\\
\enddata
\end{deluxetable*}

\begin{deluxetable}{lcccc}
\tablenum{3}
\tablecaption{Astrometric positions} 
\label{tab:astrometry}
\tablewidth{0pt}
\tablehead{
\colhead{Regions} & \colhead{RA}  & \colhead{DEC}  & \colhead{aperture radius} & \colhead{z}
}
\decimalcolnumbers
\startdata
1 & 10:27:00.375 & 17:49:00.81 & 1.25 & 0.067 \\
2 & 10:27:00.529 &  +17:49:00.57 & 1.25 & 0.066 \\
3 & 10:27:00.358 &  +17:49:02.95 & 1.25 & 0.065 \\
4 & 10:27:00.808 &  +17:49:01.41 & 1.25 & 0.066 \\
5 & 10:27:00.717 &  +17:49:00.21 & 1.25 & 0.066 \\
\enddata
\tablecomments{RA and DEC are (J2000). Aperture radius (in arcsec) used to analyze the five selected regions of the \meg\ data.}
\end{deluxetable}

\begin{deluxetable}{ccc}
\tablenum{4}
\tablecaption{Region separation
\label{tab:projsep}}
\tablewidth{0pt}
\tablehead{
\colhead{Regions} & \colhead{Distance} & \colhead{Distance}  \\
\colhead{} & \colhead{$\arcsec$} & \colhead{kpc} 
}
\decimalcolnumbers
\startdata
1-2 & 2.12 & 2.70 \\
1-3 & 1.84 & 2.35 \\
2-3 & 3.12 & 3.98 \\
\enddata
\tablecomments{The distance among the three principal nuclei in arcsec and in kpc with a z=0.066.}
\end{deluxetable}

\section{Data Analysis}
\label{section:ana}

\subsection{Stellar spectra modeling}

To model the stellar spectra, we use the {\sc pyPipe3D}\footnote{\url{https://gitlab.com/pipe3d/pyPipe3D/}} spectral fitting tool. 
This tool uses the {\sc FIT3D} pipeline for the stellar fitting that performs 
a stellar population synthesis 
\citep[SSP][]{Lacerda+2022,Sanchez16a,Sanchez16b}. 
The workflow of {\sc pyPipe3D} has been described in detail in several works 
\citep[e.g,][]{Sanchez+2022,Ibarra-Medel+2022a,Ibarra-Medel+2022b}, but we can summarize it as follows: a) Estimates the initial value of the redshift and kinematics from the central spectra. b) Performs spatial segmentation to achieve a target SNR of 50. c) Obtain per each spatial bin a first model of the stellar spectra. {\sc pyPipe3D} uses this model to obtain a gas spectrum from which it models the strong emission lines. d) {\sc pyPipe3D} subtracts the models of the strong emission lines from the original spectra and runs {\sc FIT3D} using a full stellar library to obtain the final single stellar population (SSP) synthesis. For this work, we use the gsd-156 stellar library defined in \citet{Cid-Fernandes:2013aa}. It uses 156 SSP templates that cover 39 stellar ages (from 1 Myr to 14.2 Gyr), and four metallicities ($Z/Z_{\odot} = 0.2$, 0.4, 1, and 1.5). This library uses an \citet{Salpeter+55} initial mass function (IMF). In addition, {\sc pyFIT3d} considers a \citet{Cardelli+1989} dust attenuation law for internal extinction. e) Finally, {\sc pyPipe3D} saves the results of the stellar fitting and returns a set of 2D maps of the stellar properties. The maps contain the same number of spaxels as the original data cube. 

In this study, we use the gas spectra calculated in step c) to model and calculate our emission line analysis. We did not use the {\sc pyPipe3D} analysis because we implement a more complex emission line fitting that considers a double emission line deblending and a broad component. However, we use the {\sc pyPipe3D} stellar properties returned by their SSP analysis to obtain stellar velocity dispersion and stellar masses.

\subsection{Single aperture line profiles modeling}
\label{sec:line_prof}

In this section, we present the emission line profile modeling obtained for the five regions within the \meg\ FoV. Since we are limited by the spatial resolution of the data cube, the five spectral regions were extracted with an aperture of 2.5\arcsec, i.e. two times the PSF, as we have shown in Figure~\ref{fig:regions}. 

The emission line profiles were modeled using the {\sc {IRAF}}/{\it specfit} routine \citep[][]{1994ASPC...61..437K}. This package allows us to model, at the same time, the underlying continuum and the emission lines. The line fitting was done using Gaussian profiles, for which {\it specfit} uses three input parameters: the central wavelength, the line intensity, and the FWHM. The best fit was found by minimizing the $\chi^2$ using the Marquardt algorithm for 5–10 iterations. The fits were performed in three regions, 4850-5050 \AA\ for \hb, 6200-6420 \AA\ for \oi, and 6500-6745 \AA\ for \ha. We considered a power law with a nearly flat slope for the continuum in all three regions. 

\begin{figure*}
\centering
\includegraphics[width=5cm]{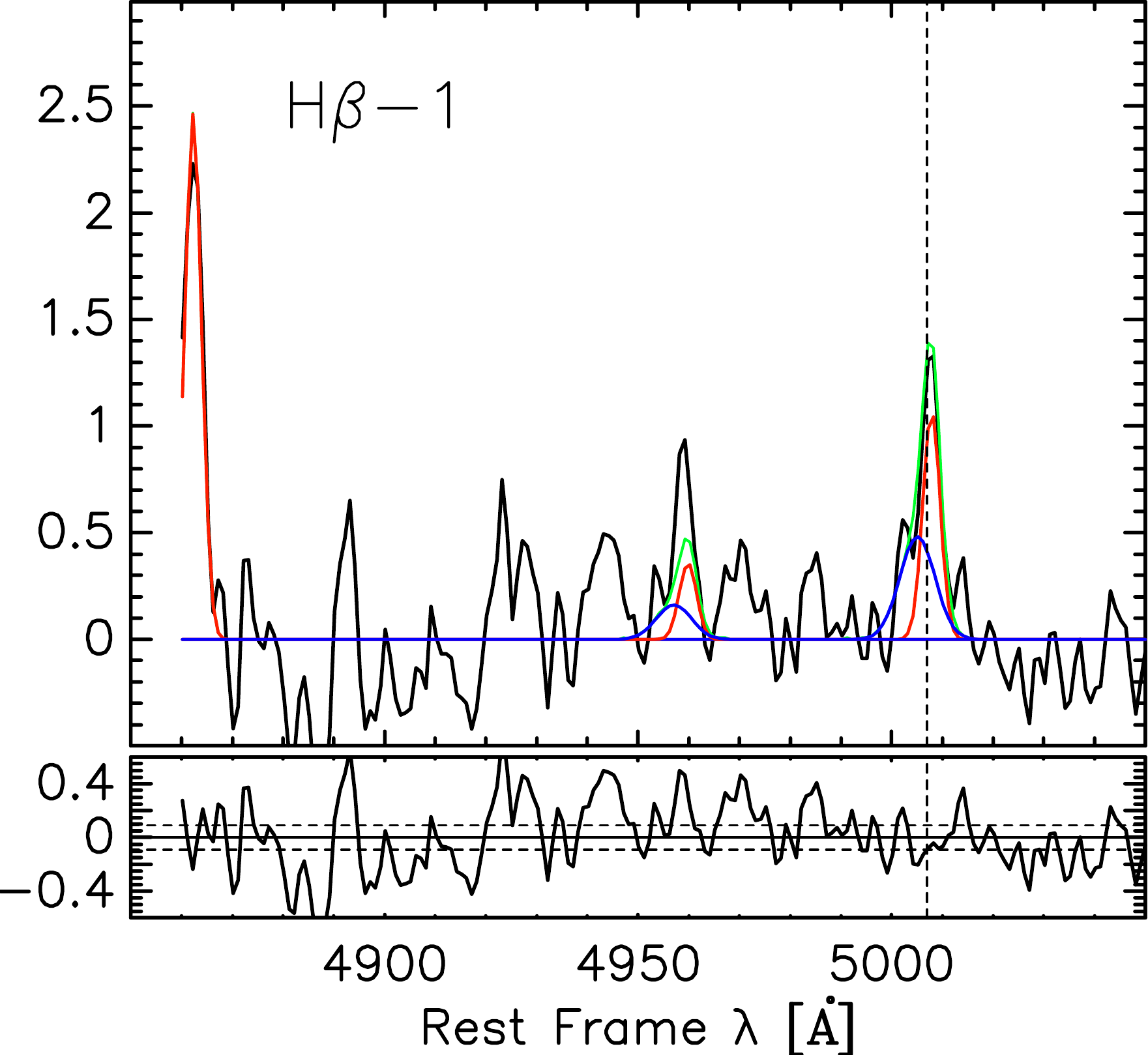}
\includegraphics[width=5cm]{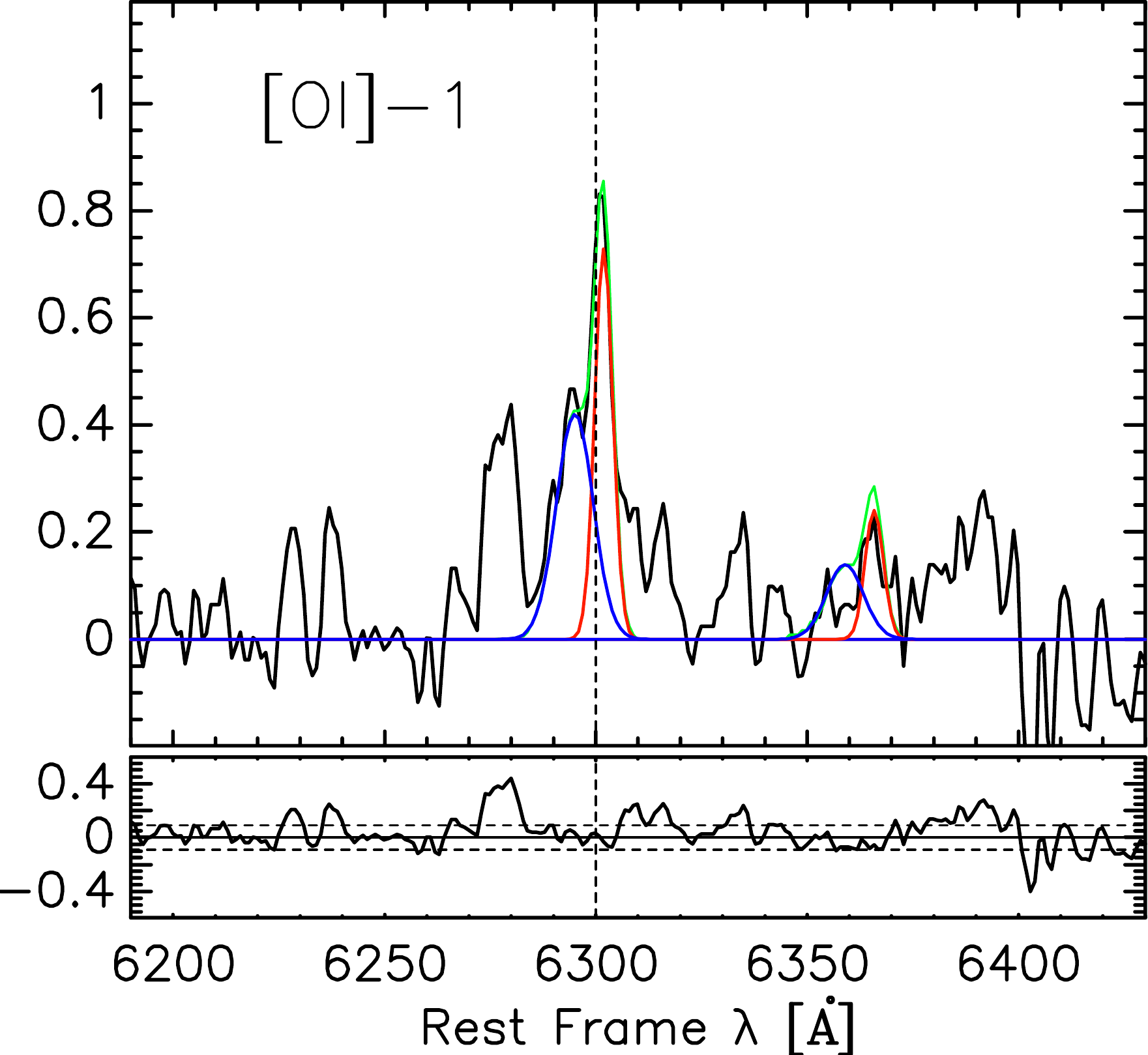}
\includegraphics[width=5cm]{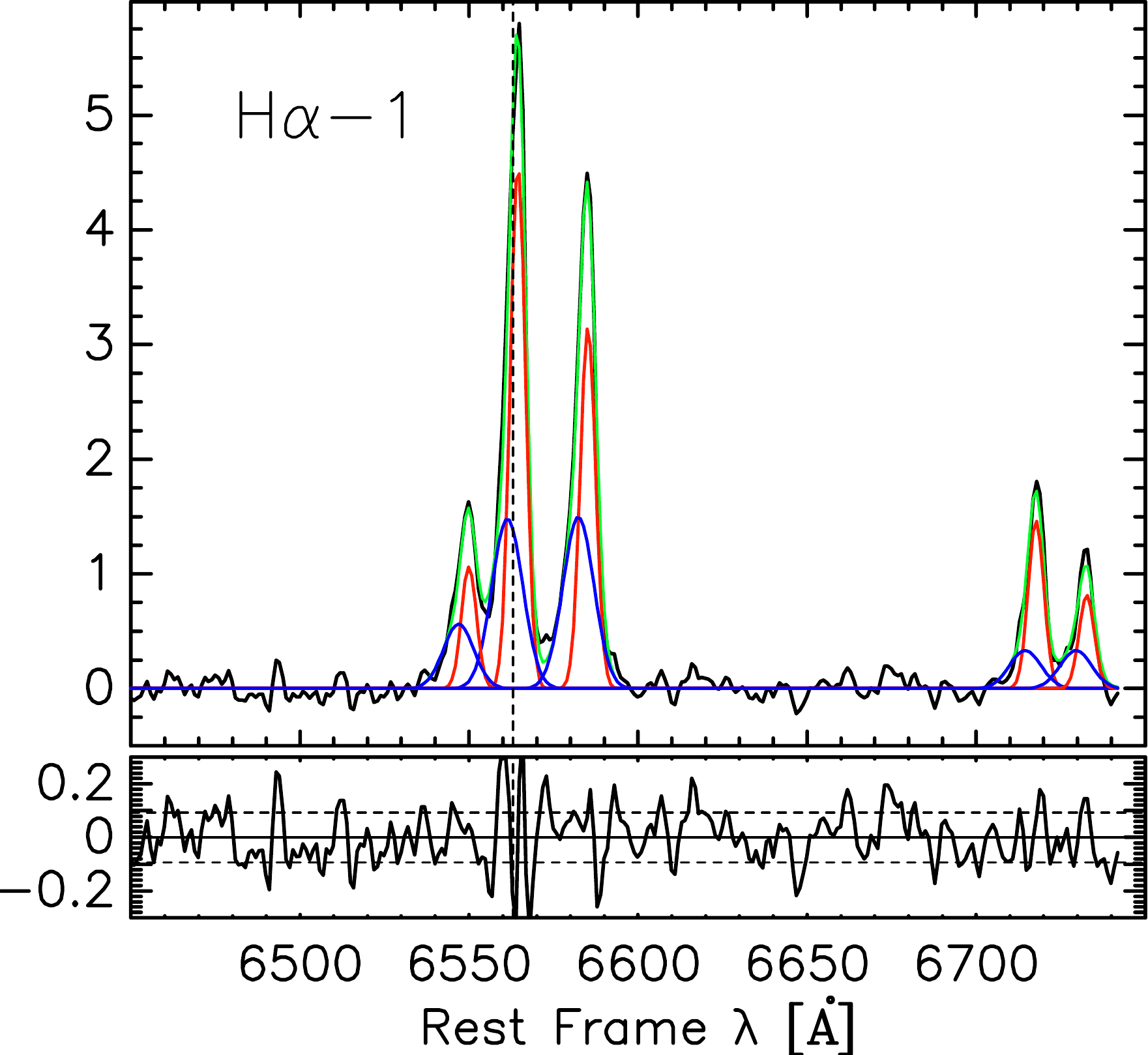}
\includegraphics[width=5cm]{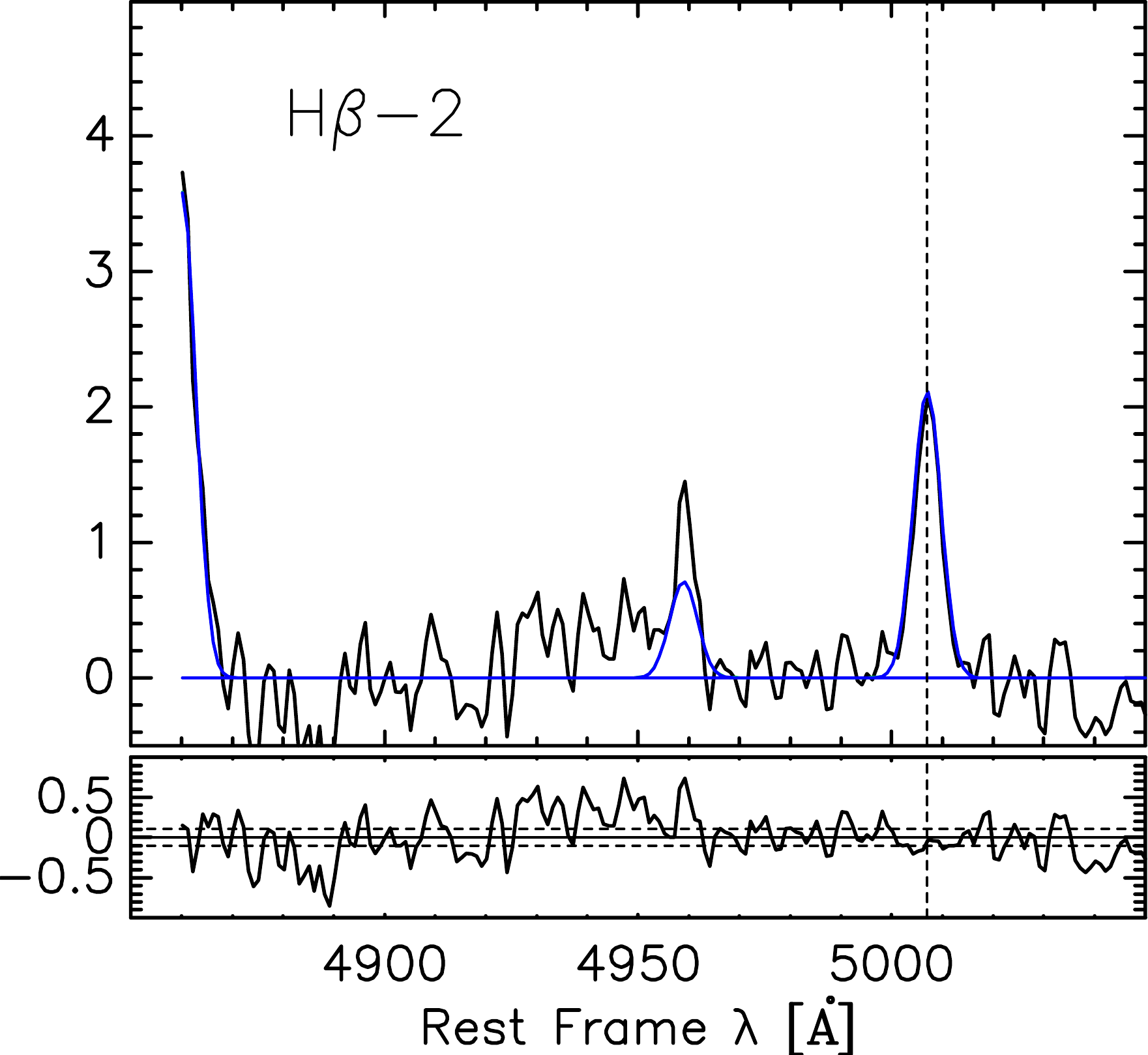}
\includegraphics[width=5cm]{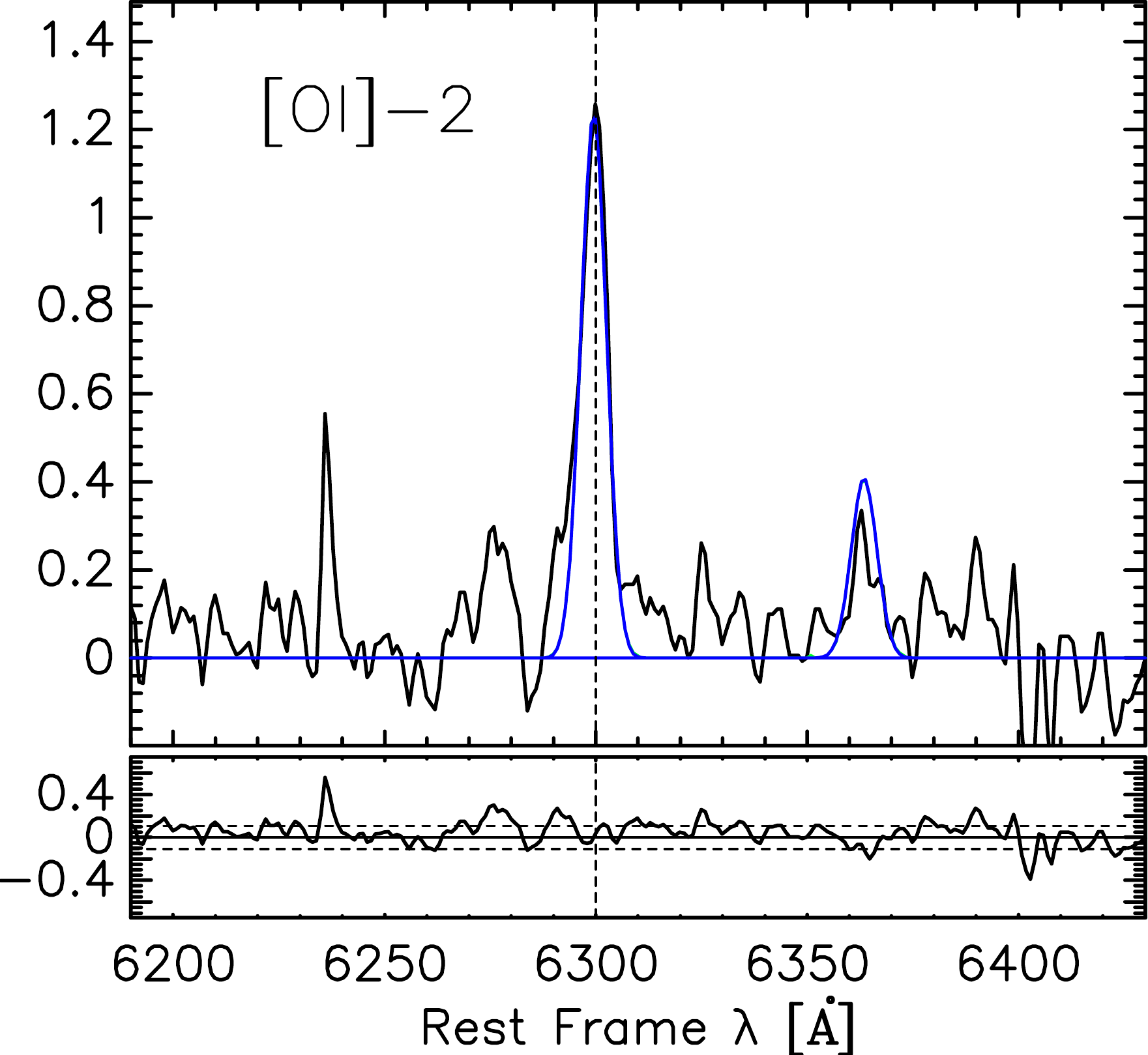}
\includegraphics[width=5cm]{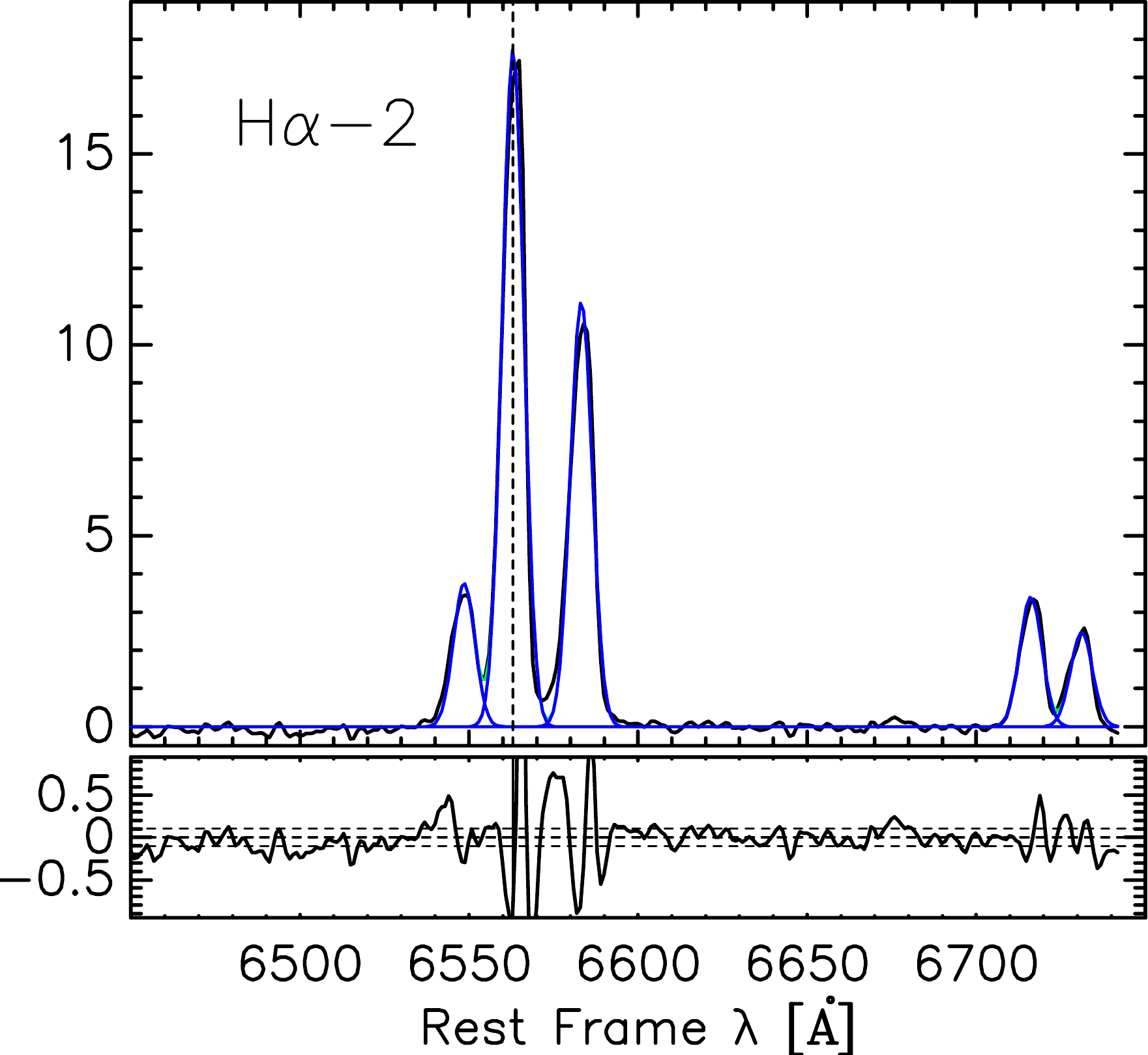}
\includegraphics[width=5cm]{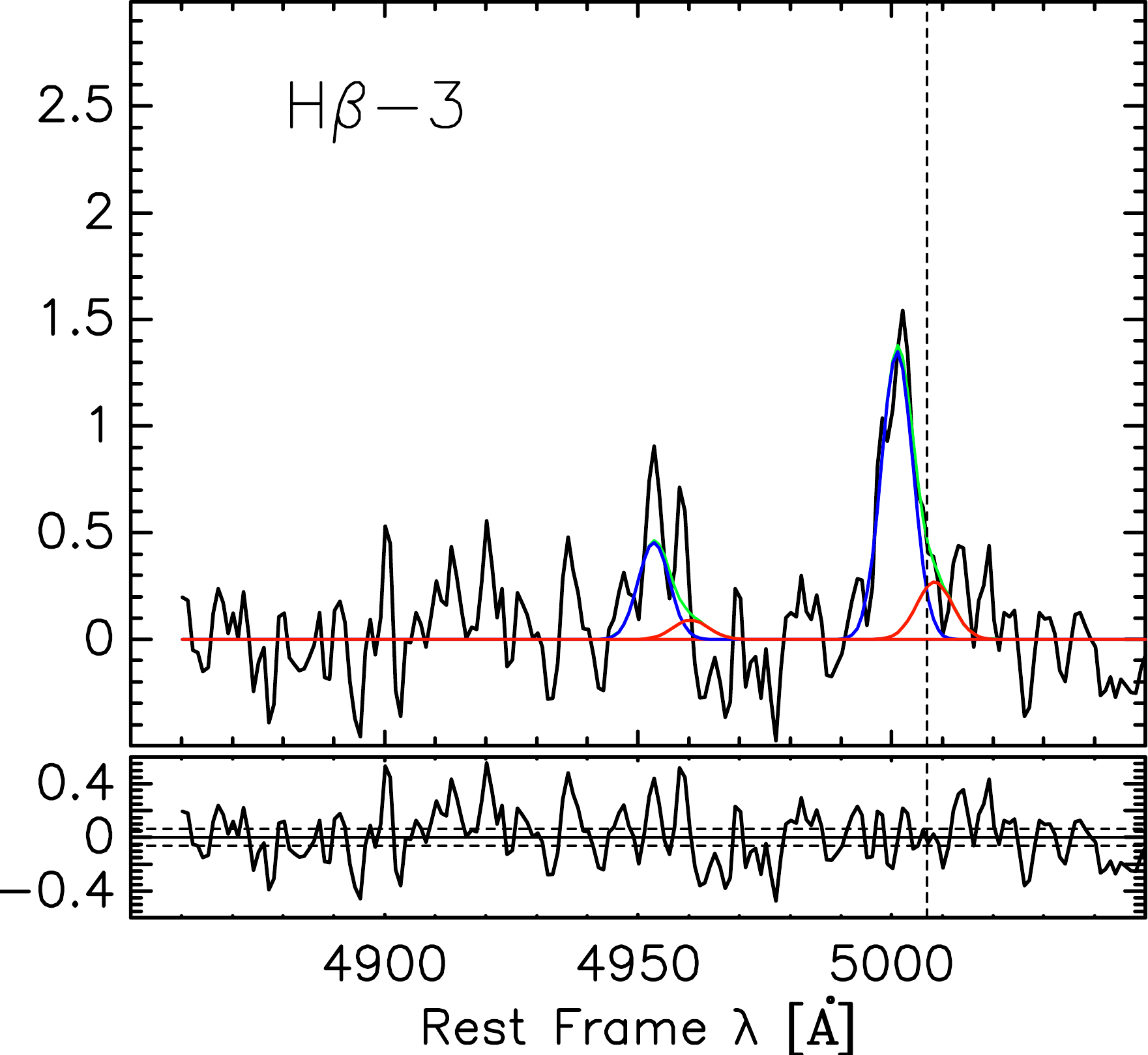}
\includegraphics[width=5cm]{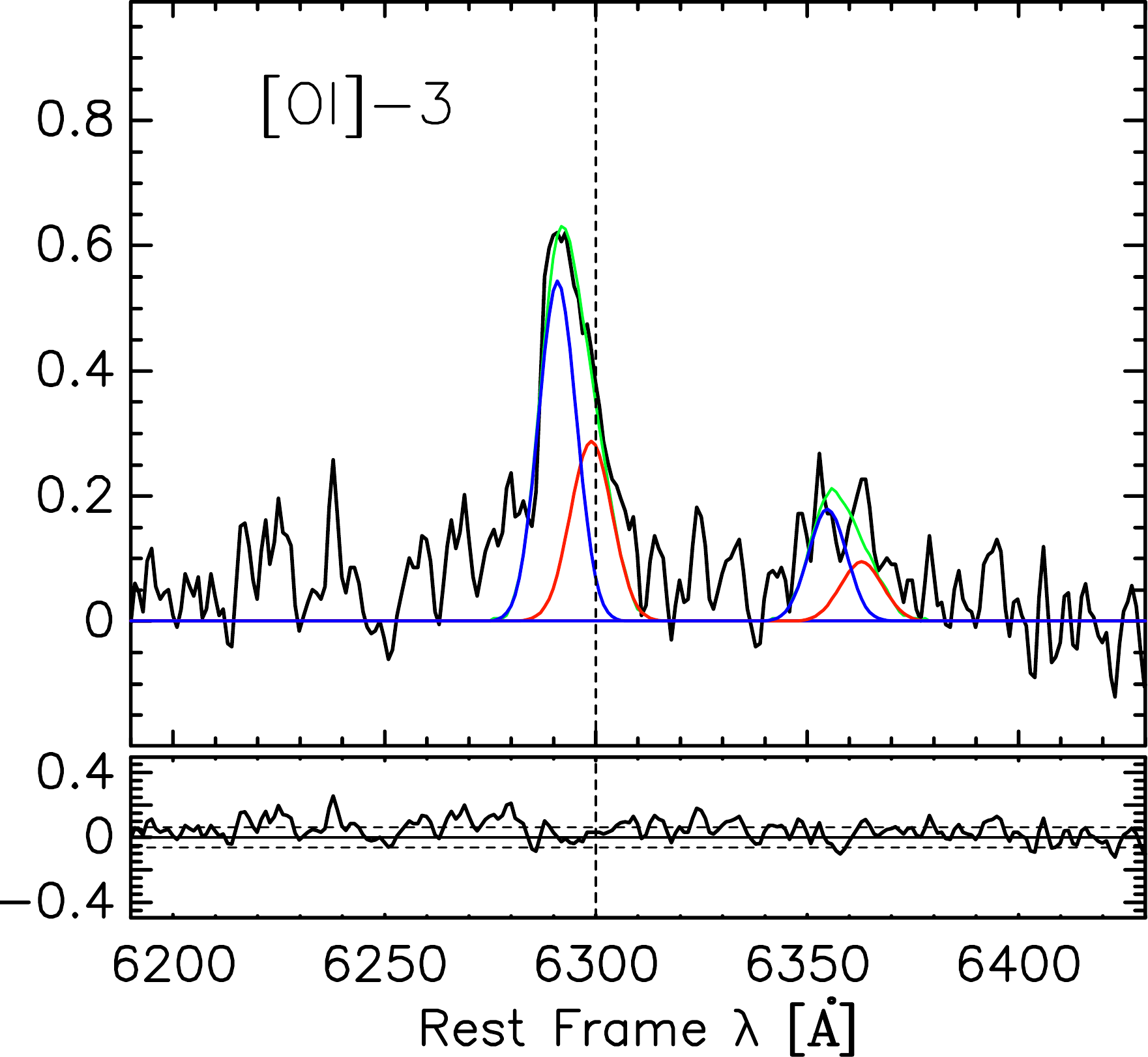}
\includegraphics[width=5cm]{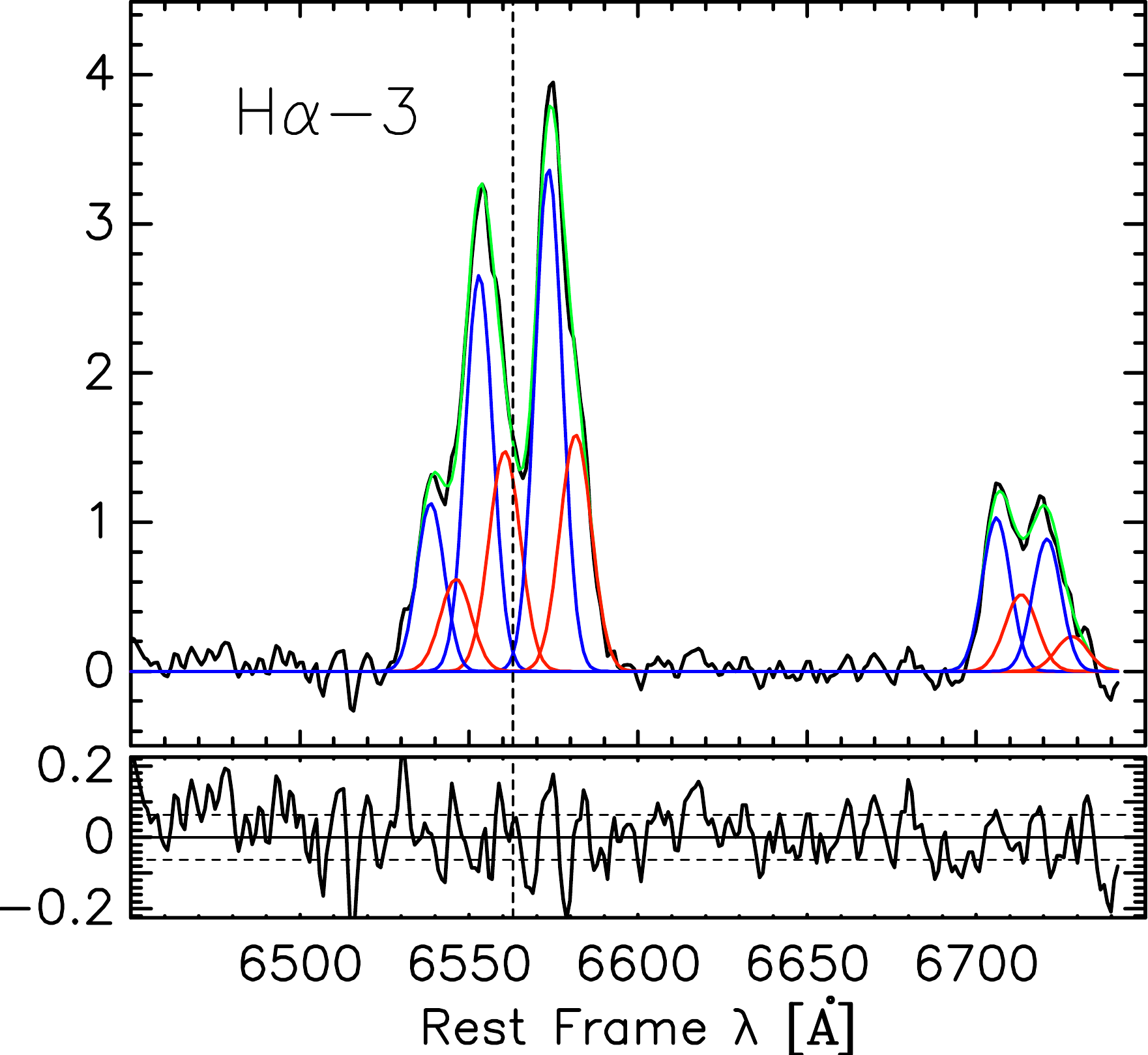}
\caption{Line profile models for Regions 1-3 with an aperture of 2.5\arcsec. Vertical axis has flux units of  $\times$10$^{-16}$erg\,s$^{-1}$\,cm$^{-2}$\AA$^{-1}$}
\label{fig:megmodelA}
\end{figure*}

We start with the line fitting in the \ha\, spectral regions because it has the most prominent lines. For all the regions, we modeled \ha, and the doublets of \nii, and \sii\, narrow components constraining the FWHM to be equal in all the lines. We assume that they originate in the same clouds and therefore share the same kinematics. The intensity of \niia\ was linked to \niib, taking into account the theoretical 3:1 ratio. The lambda shift of this doublet was also linked to being the same. The residuals obtained in the fitting of regions 1 and 3 indicate that a second Gaussian component is needed to obtain the best fit (see Figure~\ref{fig:megmodelA}).

\begin{figure*}
\centering
\includegraphics[width=5cm]{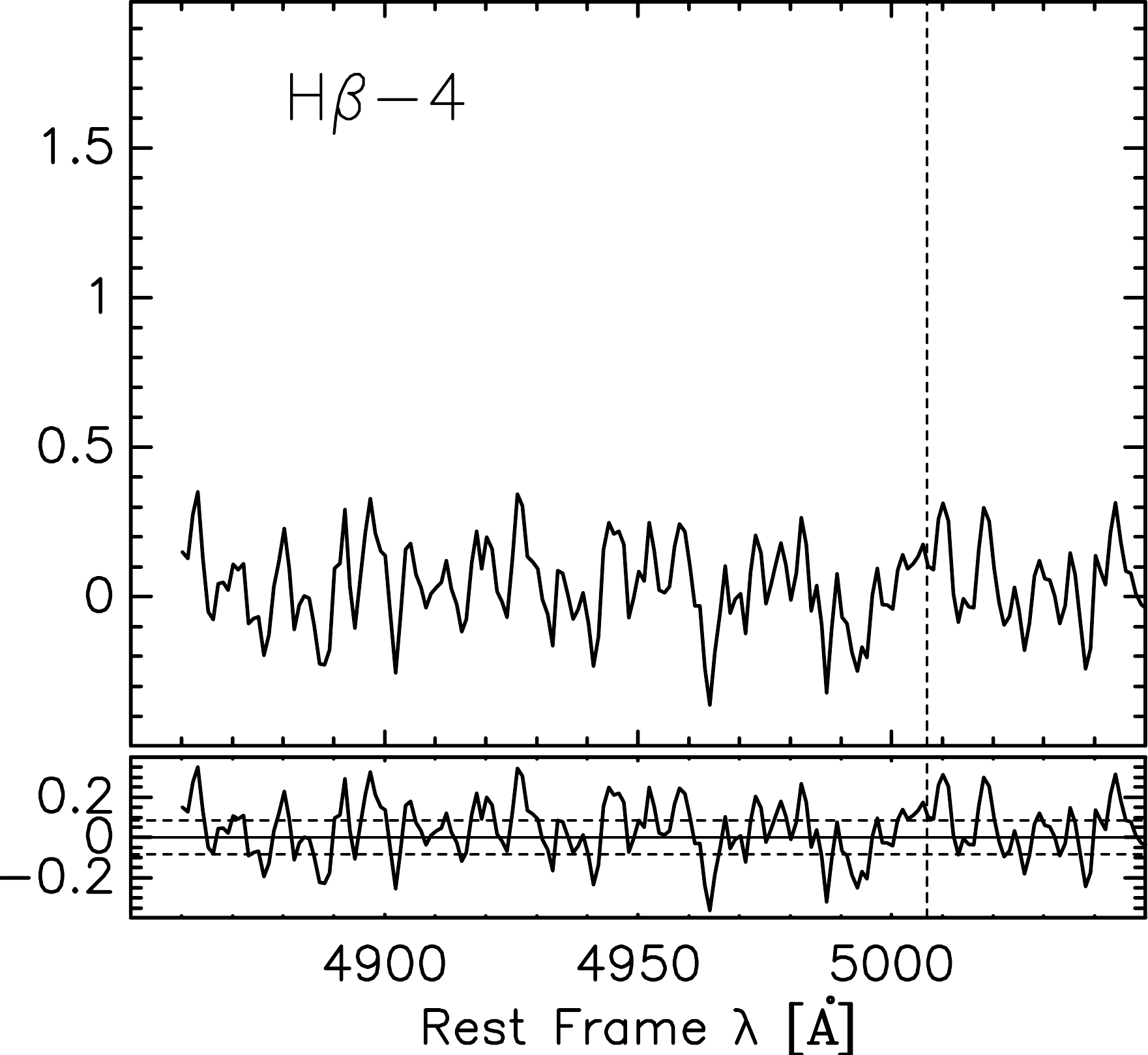}
\includegraphics[width=5cm]{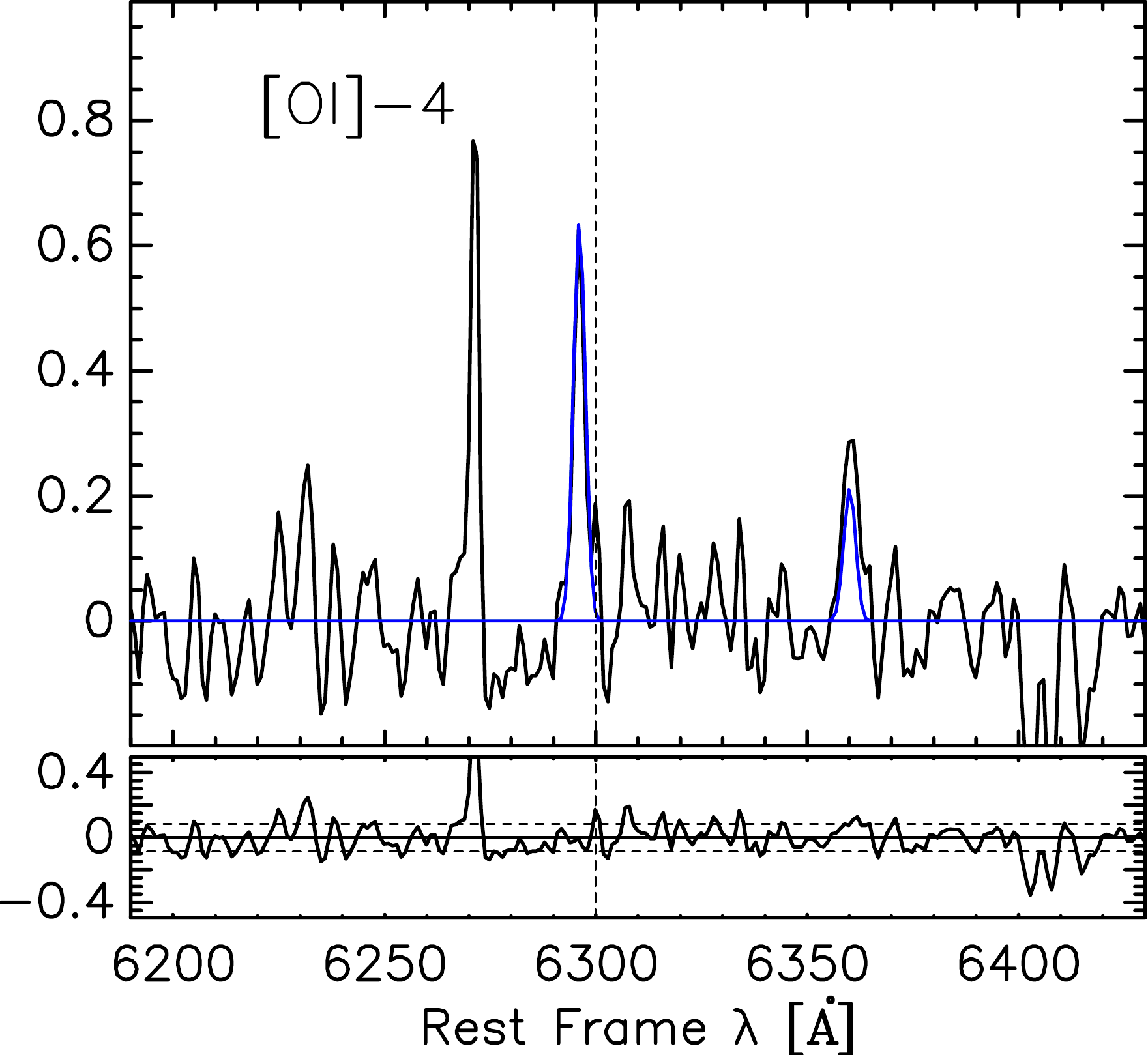}
\includegraphics[width=4.82cm]{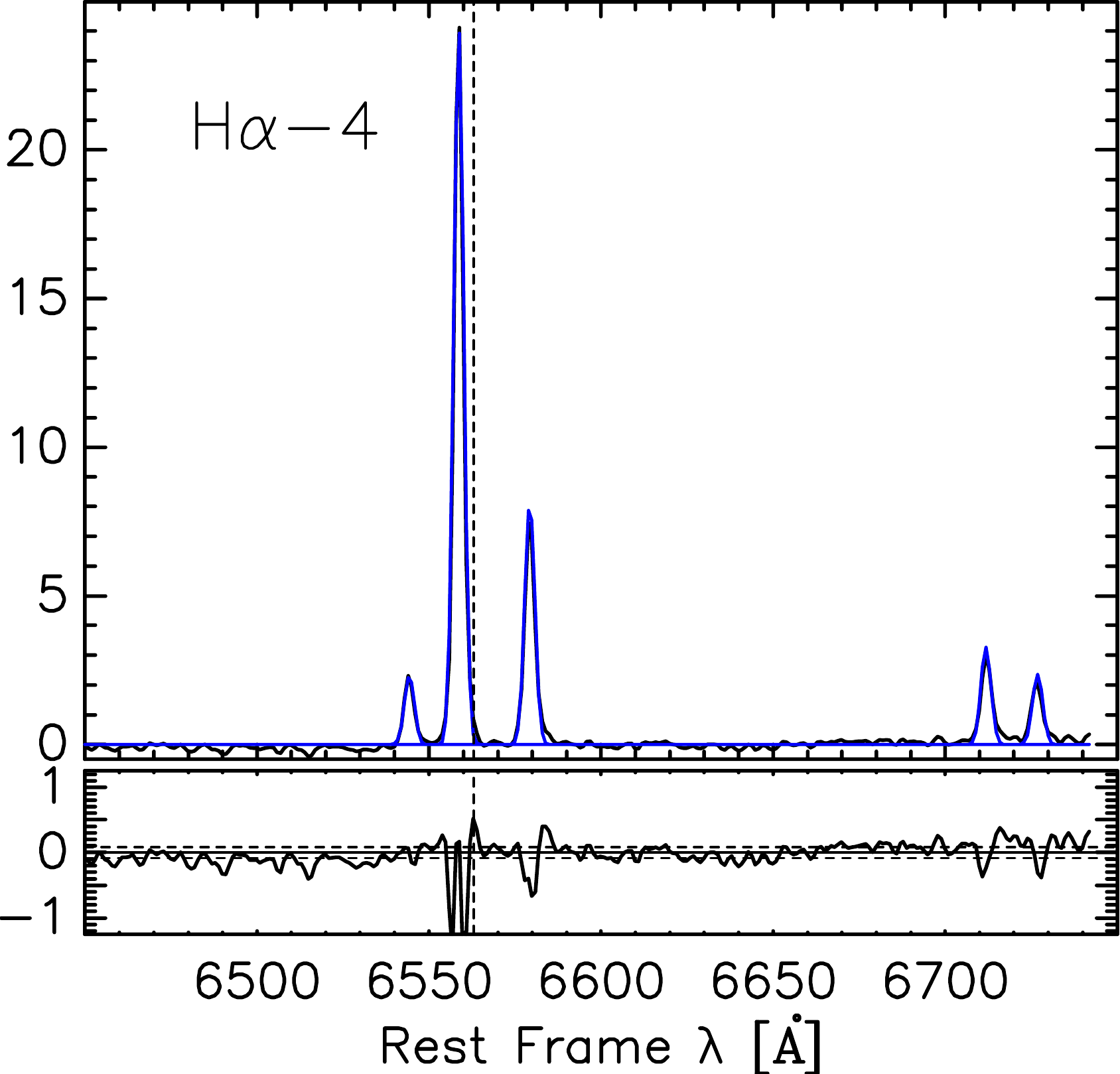}
\includegraphics[width=5cm]{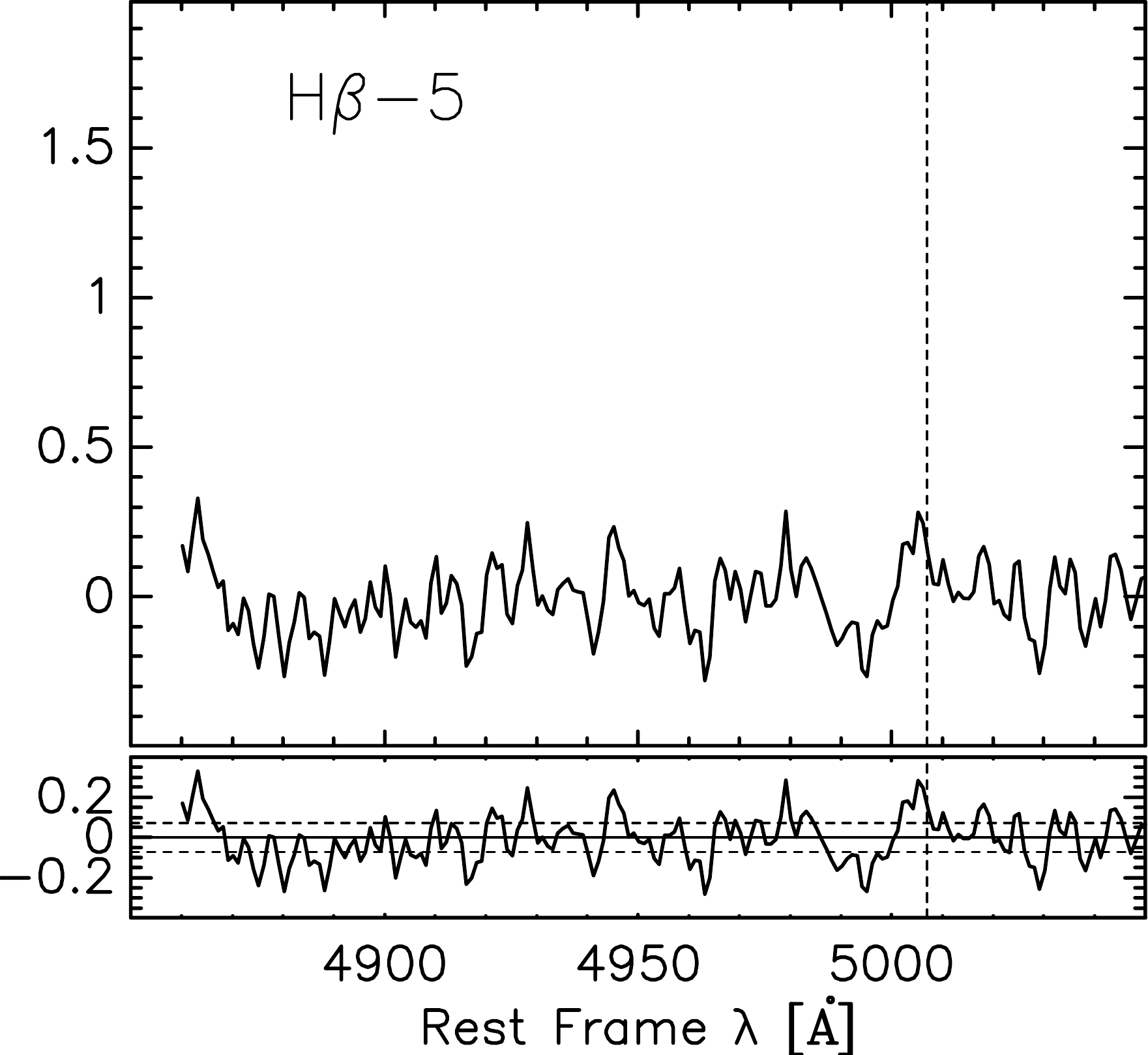}
\includegraphics[width=5cm]{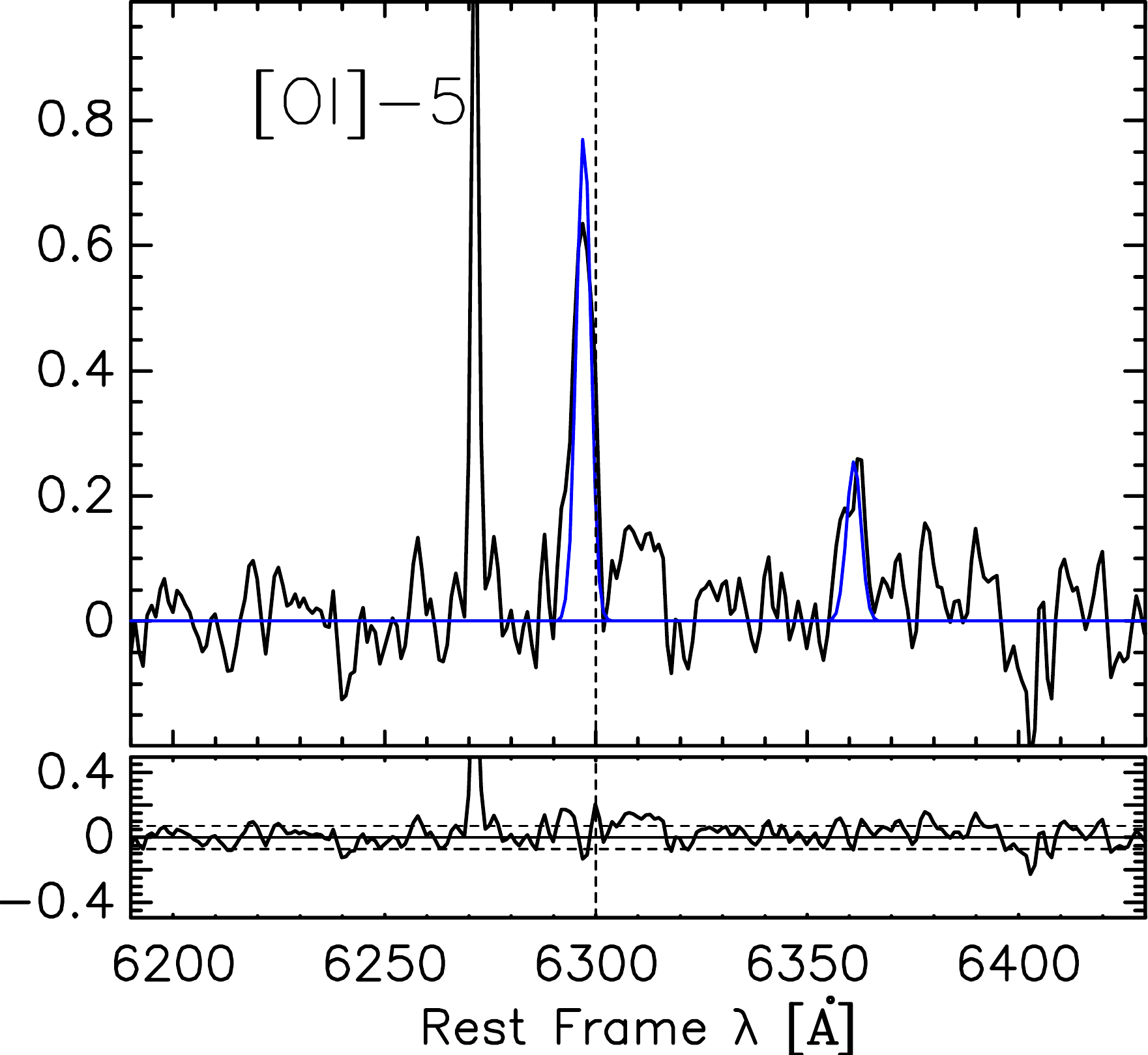}
\includegraphics[width=5cm]{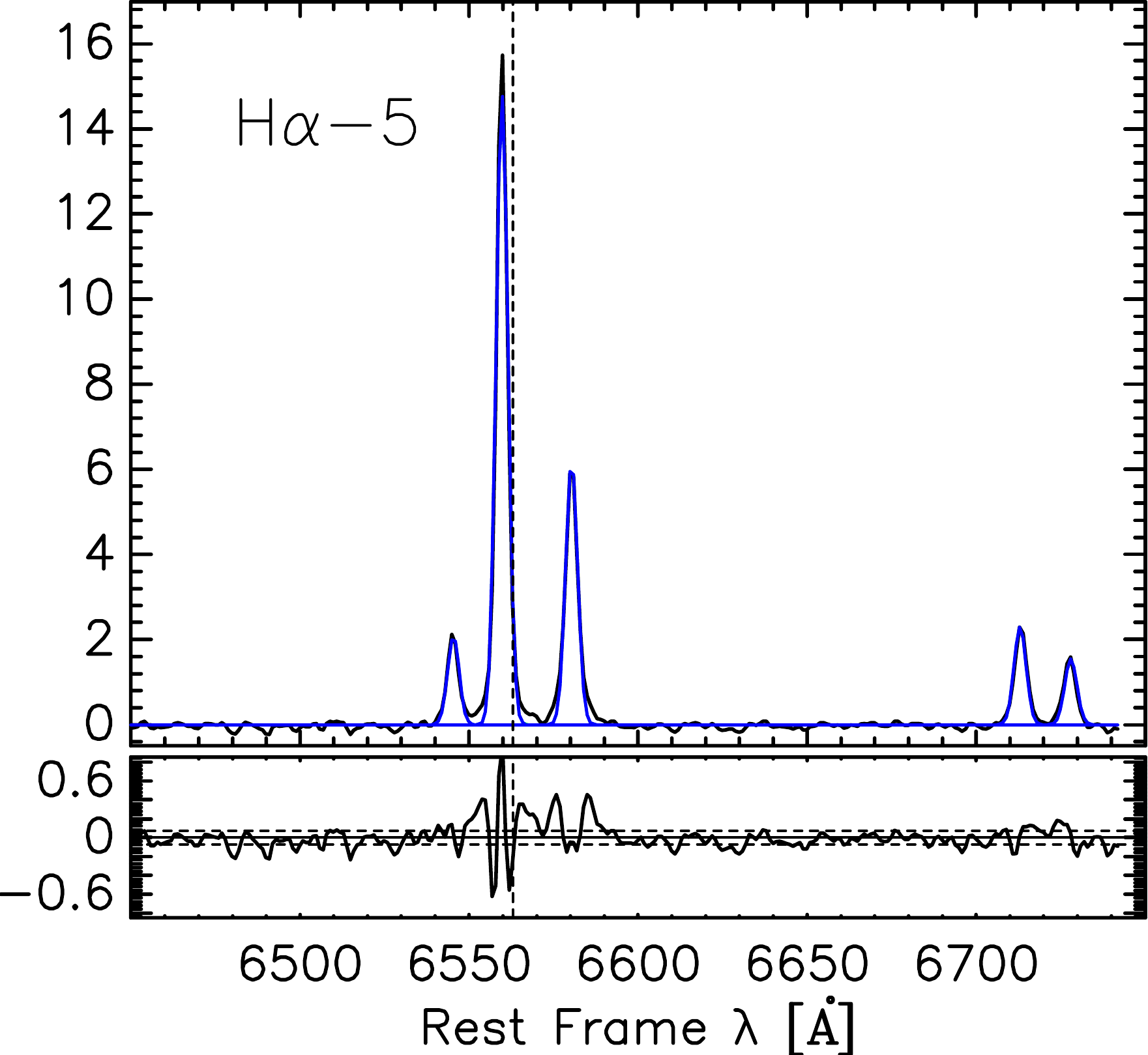}
\caption{Line models for Regions 4 - 5 with an aperture of 
2.5\arcsec. The vertical axis has flux units of  
$\times$10$^{-16}$\,erg\,s$^{-1}$\,cm$^{-2}$\AA$^{-1}$}
\label{fig:megmodelB}
\end{figure*}

Therefore, for regions 1 and 3, the constraints on the FWHM were considered for each set of lines (blue and redshifted components). \objedos\ instead, was modeled with a single Gaussian component.

The \hb\, and \oi\, spectral regions were fitted with the same components as their respective \ha\ regions, with the doublets of \oi\, and \oiii\, fitted similarly to \nii. Only in regions 1, 2, and 3 \oiii\, is visible. In regions 1 and 2, we can see hints of \hb\, emission, for which we attempted to fit a Gaussian component with the same FWHM as \oiii. However, the SNR is not enough to achieve a good fit, so these lines were not analyzed. It is worth mentioning that in the \meg\, spectra of these three regions, the H$\beta$ emission line is dominated by noise due to the low exposure time of the VPH480 (see left panels of Figure~\ref{fig:megmodelA}). The line profile modeling results for regions 4 and 5 are shown in Figure~\ref{fig:megmodelB}. Both regions were fitted with single Gaussian components and are found to be star formation (SF) regions. The FWHM of the emission line profiles obtained for the five regions is presented in Table~\ref{table:main}. 

\subsection{2D spatially resolved emission line profile modeling}
\label{2d_linefit}

\begin{figure}
\centering
\includegraphics[width=1.0\columnwidth]{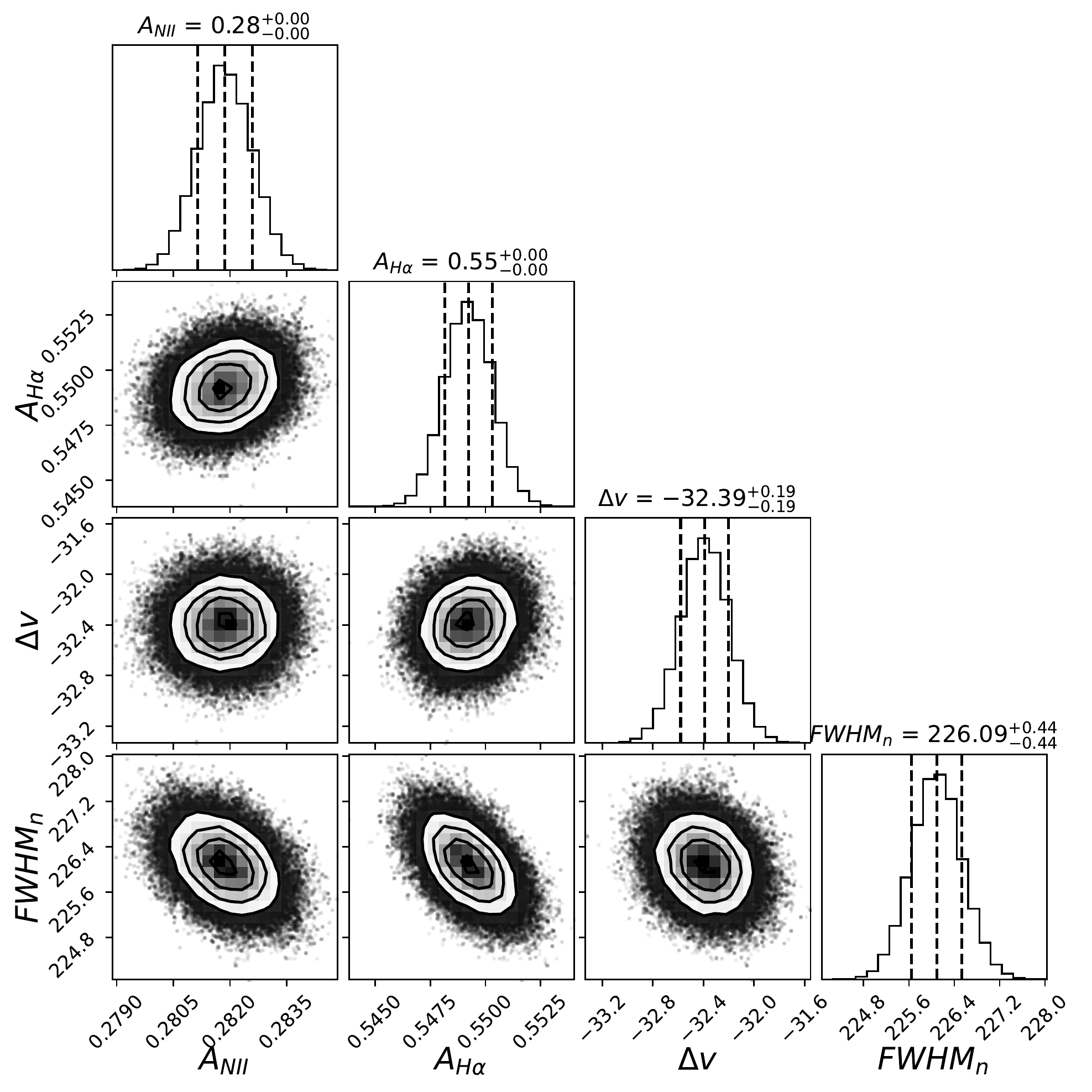}
\includegraphics[width=1.0\columnwidth]{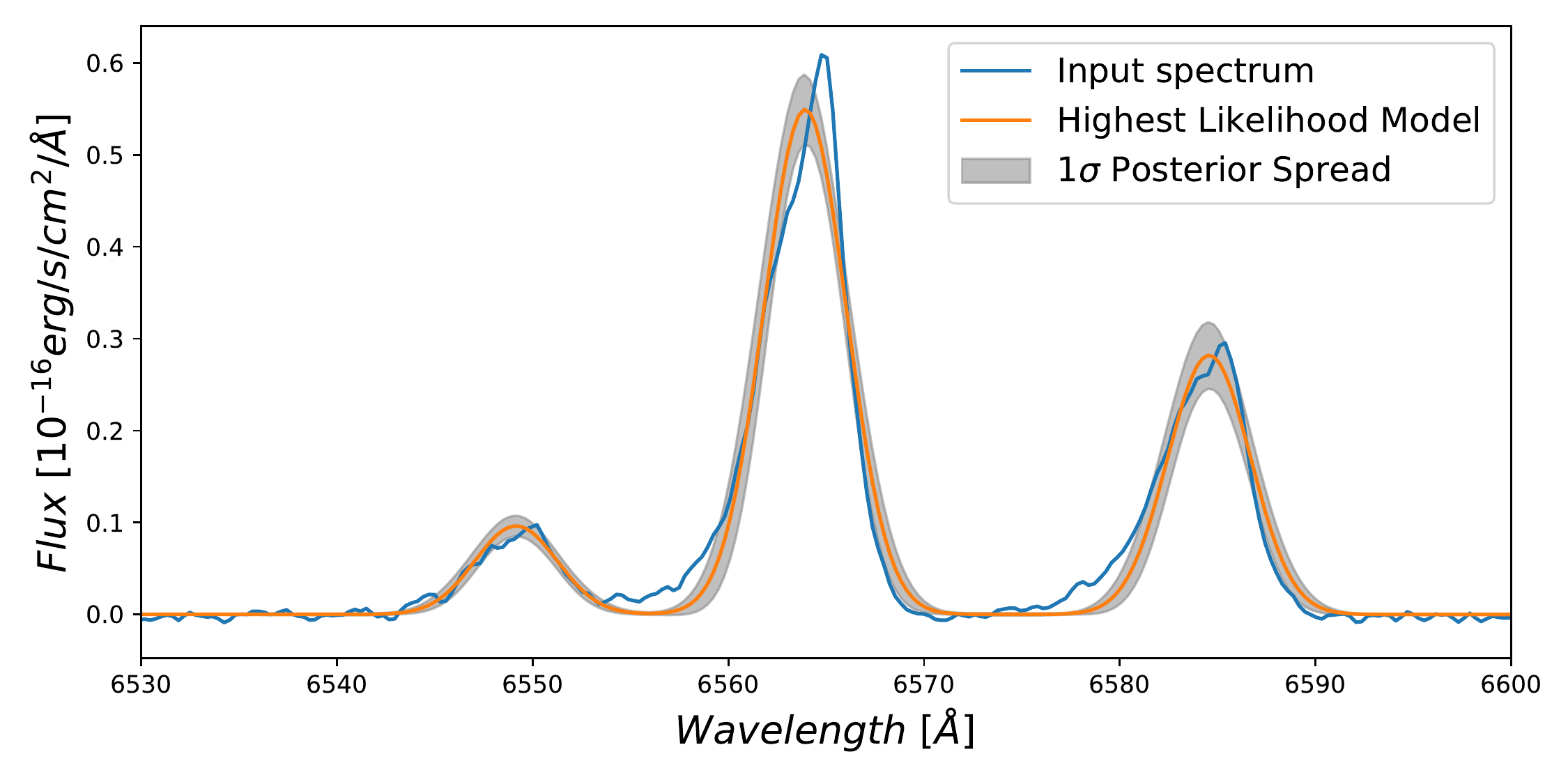}
\caption{Example of a single spaxel line fitting model of \ha \ and \nii \ using the MCMC method and a Gaussian profile. In the upper panel, we show the confidence plots of four out of seven free parameters of our fit: the amplitudes of \ha, \nii 6585, the velocity shift $\Delta v_n$ of the narrow components, and the FWHM of the narrow components. We did not find the presence of a broad component. Lower panel: We plot the input spectra of the \ha \ region  with the best-fit model, the shaded region shows the dispersion of our posterior line fit models within $1-\sigma$ confidence.}
\label{spec_fit_mcmc}
\end{figure}

To exploit the full advantage of the IFS spatially resolved data, we implement an MCMC modeling of the \nii 6549, \nii 6585, and \ha \ emission lines spaxel by spaxel. In the previous section, we applied meticulous and detailed modeling of the strong emission lines for regions 1,2,3,4, and 5. 
In this section, we applied a simplified fit that only considers one single narrow and broad component for the \ha \ region. We use the \citet{emcee} implementation of the Affine Invariant MCMC Ensemble Sampler with the use of the Python package 
{\sc emcee}\footnote{https://emcee.readthedocs.io/en/stable/} to model the \ha \ region. The space parameter that we explore is the amplitudes of \ha, \nii 6585, the velocity shifts $\Delta v_n$ of the narrow components, the FWHM of the narrow components, the FWHM of the broad component, the velocity shift $\Delta v_b$ of the broad component, and the amplitude of the broad component. The model fixes the relative velocity among \ha, \nii 6549, and \nii 6585 with the values reported in \citet{VandenBerk+2001}, and fixes the amplitude ratio of \nii 6585 / \nii 6549 to 2.93. The velocity shift of the \ha \ broad component is left free. In Figure~\ref{spec_fit_mcmc}, we show an example of the MCMC \ha \ region fitting in the central spaxel of our data cube. With this analysis, we can reconstruct the 2D spatially resolved maps of the kinematic components of the narrow lines.

\section{Results}
\label{section:res}

\subsection{Velocity maps}\label{sec:kinematic}

\begin{figure*}[p]
  \sbox0{\begin{tabular}{@{}cc@{}}
  \includegraphics[width=1.15\textwidth]{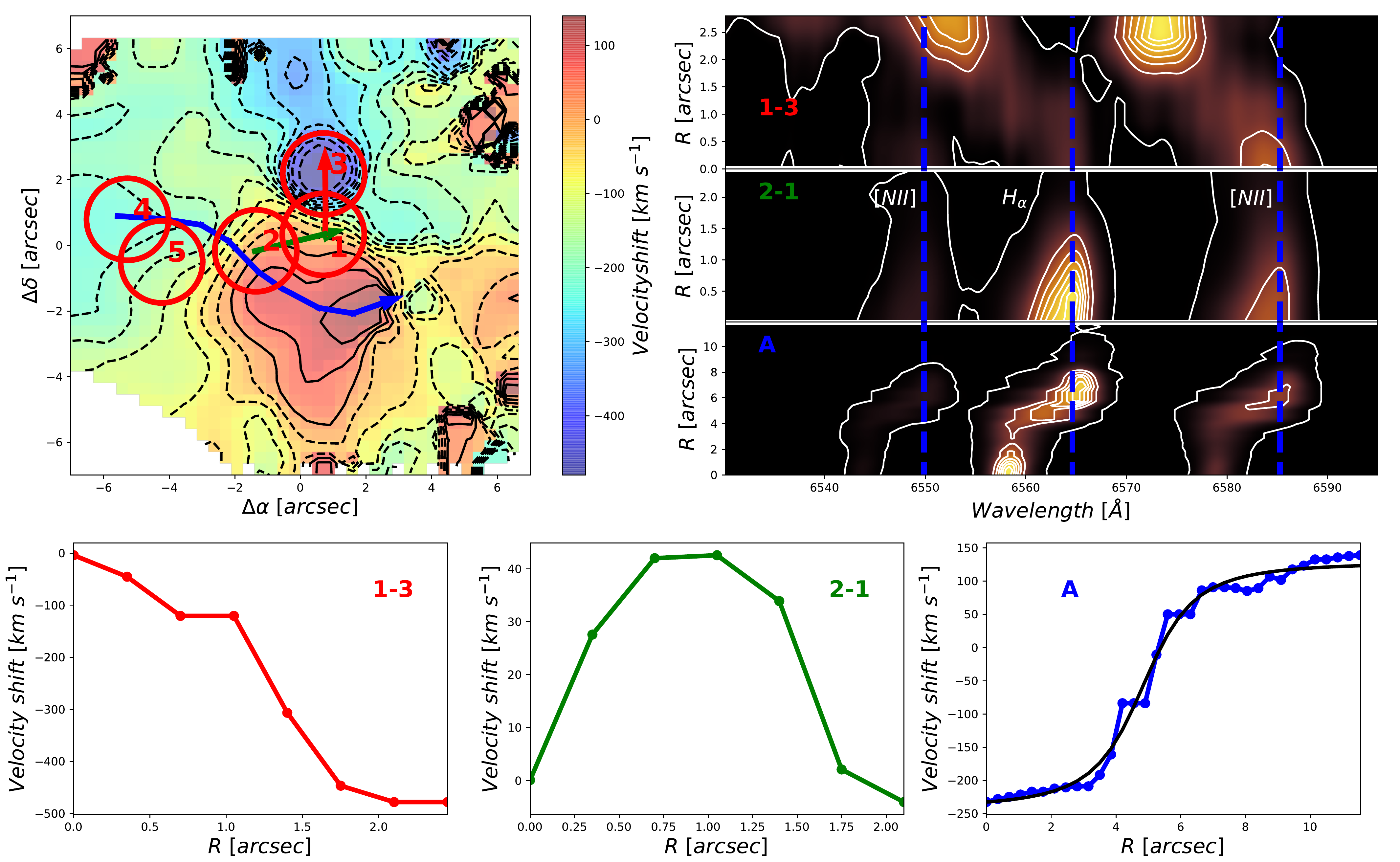}
  \end{tabular}}
  \rotatebox{90}{\begin{minipage}[c][\textwidth][c]{\wd0}
    \usebox0
    \caption{Upper Left panel: Velocity map of the narrow emission lines of \ha, with the path 1-3 (red arrow), 2-1 (green arrow), and A (blue arrow) showing the paths of the explored velocity profiles. Upper Right panel: 2D emission spectra of the \ha\ region within the segments 1-3, 2-1, and A. The dashed blue lines represent the wavelengths of \ha\ and [NII] emission lines in the \J1027\ (region 1) rest frame. Lower panels: Velocity profiles along paths 1-3, 2-1, and A. For path A, the black solid line represents the best velocity profile fit using the \citet{Bertola+1991} parameterization.}
    \label{havelmap}
  \end{minipage}}
\end{figure*}

From the spatially resolved line profile fitting, we are able to obtain not only the velocity shifts of the narrow lines for our five regions but also we are able to reconstruct the full line-of-sight (l.o.s) kinematic field of the narrow component within the \meg\ FoV. In Figure~\ref{havelmap}, we show the results of the spatially resolved kinematics. Here, we also did not find any evidence of the existence of a broad component in the three nuclei.  The velocity map of the narrow components shows that part of the gas of \objetres\ is blueshifted with a maximum velocity of -500\,\kms. 
In the case of object \objedos, iso velocities curves trace the rotation from a gaseous disk. In addition, the full velocity field shows the presence of an extended emission towards the SE and W up to $\sim$6\arcsec, which corresponds to $\sim$7.6\,kpc, and less extended towards the NE $\sim$4\arcsec or $\sim$5.1\,kpc. 

\subsection{Velocity Shifts} 
\label{subsection:rot}

Using the resolved velocity field, we explore the velocity profiles moving from \objedos\ (region 2) to \J1027 (region 1), denoted as path 2-1, and a green arrow in Figure \ref{havelmap}. We also explore the velocity profiles moving from object \J1027\ (region 1) to \objetres\ (region 3), path 1-3 marked with a red arrow in Figure \ref{havelmap}. The velocity profile along path 1-3 clearly shows a blue shift of $\approx$-500\,\kms for object \objetres\ with respect to \J1027\ and \objedos. The velocity profile in 2–1 shows a velocity increment from zero (we set the velocity rest frame to the center of \objedos) to $\approx\,$42\,\kms at the middle of the path and then drops to -3\,\kms. This shape shows the velocity changes of the observed rotation of the gaseous disk in object \objedos\ along path 2-1. 

In addition, in the upper right panel of Figure~\ref{havelmap}, we show the actual line shifts along the paths 1-3, 2-1, and path A (blue arrow in Figure \ref{havelmap}) for \ha, \niia\ and \niib. The intensity map on that panel represents the total flux intensity of the lines in the spectral range 6530-6600 \AA\ along the paths. We also overplotted  the rest-frame wavelengths of \ha, \niia\, and \niib\ centered in \J1027, as a segmented blue line. For paths 1-3 (from region 1 to region 3), we observe how the emission lines start to have a blue shift at $\approx$\,1.2\arcsec from the center of \J1027. At $\approx$\,1.7\arcsec, the emission lines reach their final shift ($\approx$\,-500\,\kms). For paths 2-1, we observe a small shift, but we show that the principal emission line flux comes from region 2. And finally, in path A, we clearly observe how the emission lines have a shift from blue to red along path A, which is centered in region 2.

We now explore in more detail the shifts along path A, which traces the rotation of the gaseous disk in the object \objedos. We measure the velocity change along path A (blue arrow in Figure \ref{havelmap}). Path A pass through the object \objedos, and it defines the path of the maximum velocity gradient: it passes perpendicularly through all iso velocity contours. The velocity profile along path A (right lower panel of Figure \ref{havelmap}) shows a clear rotation curve that has a dynamical center within the aperture 2 where \objedos\ lies. It has a minimum velocity of -230\,\kms, and a maximum velocity of 140\,\kms. To characterize this rotation curve, we use the parameterization of \citet{Bertola+1991} to obtain the value of $V_{max}$, as discussed in \citet{Aquino+2018}. We show this parameterization as a black solid line of the right lower panel of Figure \ref{havelmap}. We get an $V_{max}$ of 128.6\,$\pm$6\,\kms, with a systemic velocity of -57\,$\pm$3\,\kms, and a rotation curve amplitude of 185.6\,$\pm$6\,\kms.

\subsection{Stellar velocity dispersion and stellar masses}
\label{stellarmass}

Using the outputs of {\sc pyPipe3D}, we recover the SSP maps of the velocity dispersion ($\sigma_{\star}$), the total stellar flux, and the stellar mass within the FoV of \meg. To calculate the total velocity $\sigma_{\star}$ within all apertures, we get the quadratic mean weighted by the stellar flux ($F_{ssp}$) of all the spaxels within each region, defined as:

\begin{displaymath}
\sigma_{\star}=\left (\frac{\sum_{i,j}\sigma_{\star,i,j}^2\times F_{ssp,i,j}}{\sum_{i,j}F_{ssp,i,j}}\right )^{1/2},
\end{displaymath}
the indexes $i,j$ are the indexes within the apertures of our regions. To obtain the total stellar mass, we simply get the total stellar mass contribution of all the spaxels within each region. We present the final values for the five regions in Table~\ref{table:main}.

\subsection{HST images}
\label{section:HST_image}
We present the analysis of the high-resolution HST $U$- and $Y$-band imaging of the target obtained by Program GO 13112 (PI: Liu). Details on these observations are given in Table~\ref{tab:hst}. For our study, we use the final reduced exposures taken with the Wide Field Camera 3 (WFC3) in the  UV filter F336W (i.e., $U$ band) and in the IR filter F105W. In addition, the WFC3 (i.e., $Y$ band) instrument handbook\footnote{from the HST user documentation: \url{https://hst-docs.stsci.edu/wfc3ihb}} indicates that the FWHM PSF values are on the order of $0.13\arcsec$ for the F105W and $0.08\arcsec$ for the F336W. Those exposures provide the necessary spatial resolution to resolve any substructure within our five regions.

\begin{deluxetable}{cccc}
\tablenum{5}
\tablecaption{HST data
\label{tab:hst}}
\tablewidth{0pt}
\tablehead{
\colhead{Filter} & \colhead{exptime} & \colhead{$\lambda$} & FWHM PSF \\
\colhead{} & \colhead{$s$} & \colhead{\AA} & $\arcsec$ 
}
\decimalcolnumbers
\startdata
F105W & 238.704 & 10551.047 & 0.13 \\
F336W & 2154.000 & 3354.656 & 0.08\\
\enddata
\tablecomments{\J1027\ HST data were obtained with WFC3 in 2013 03 27. PSF values were obtained from the WFC3 
HST user documentation.}
\end{deluxetable}

\begin{figure*}
\centering

\includegraphics[width=1.0\columnwidth]{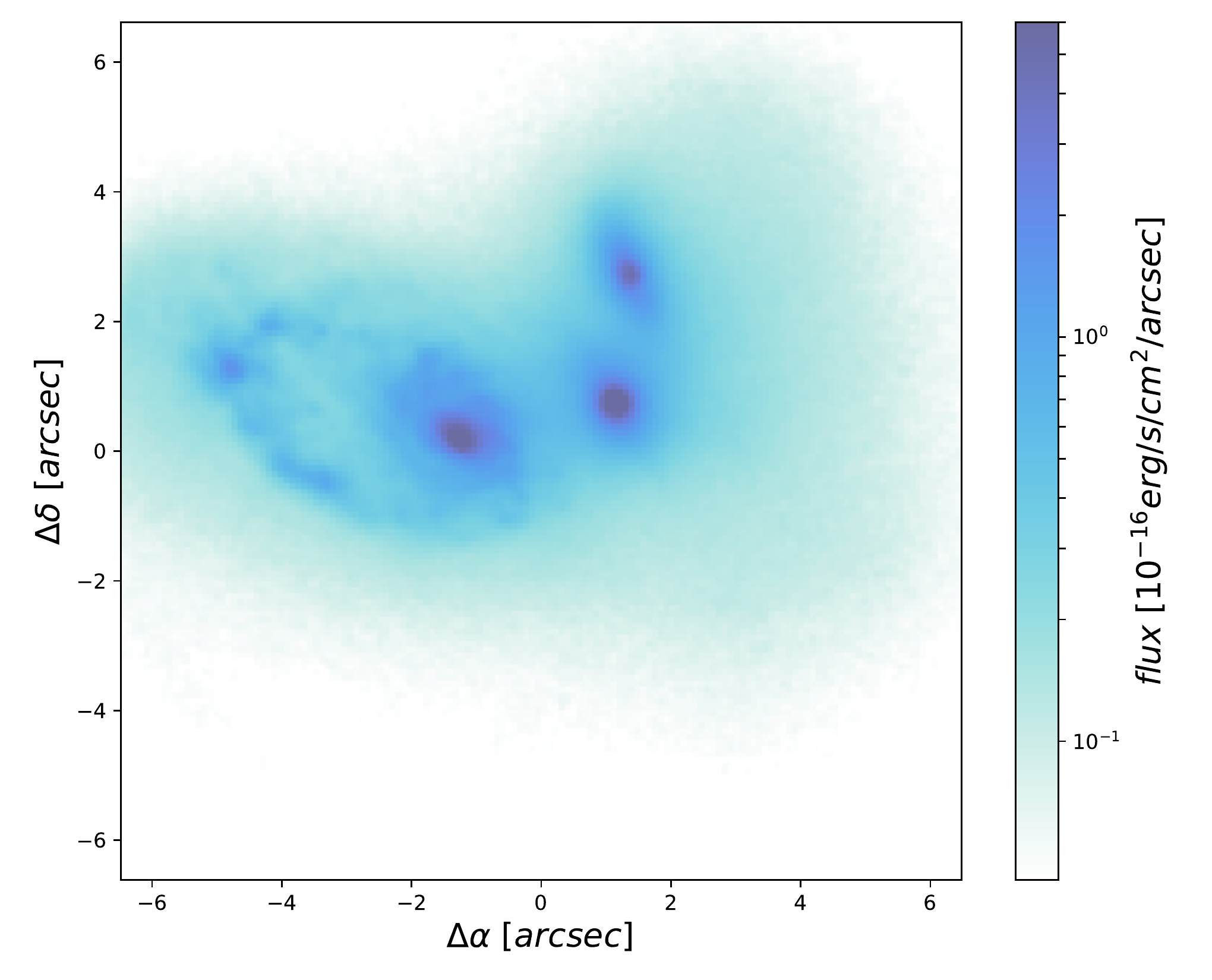}
\includegraphics[width=1.0\columnwidth]{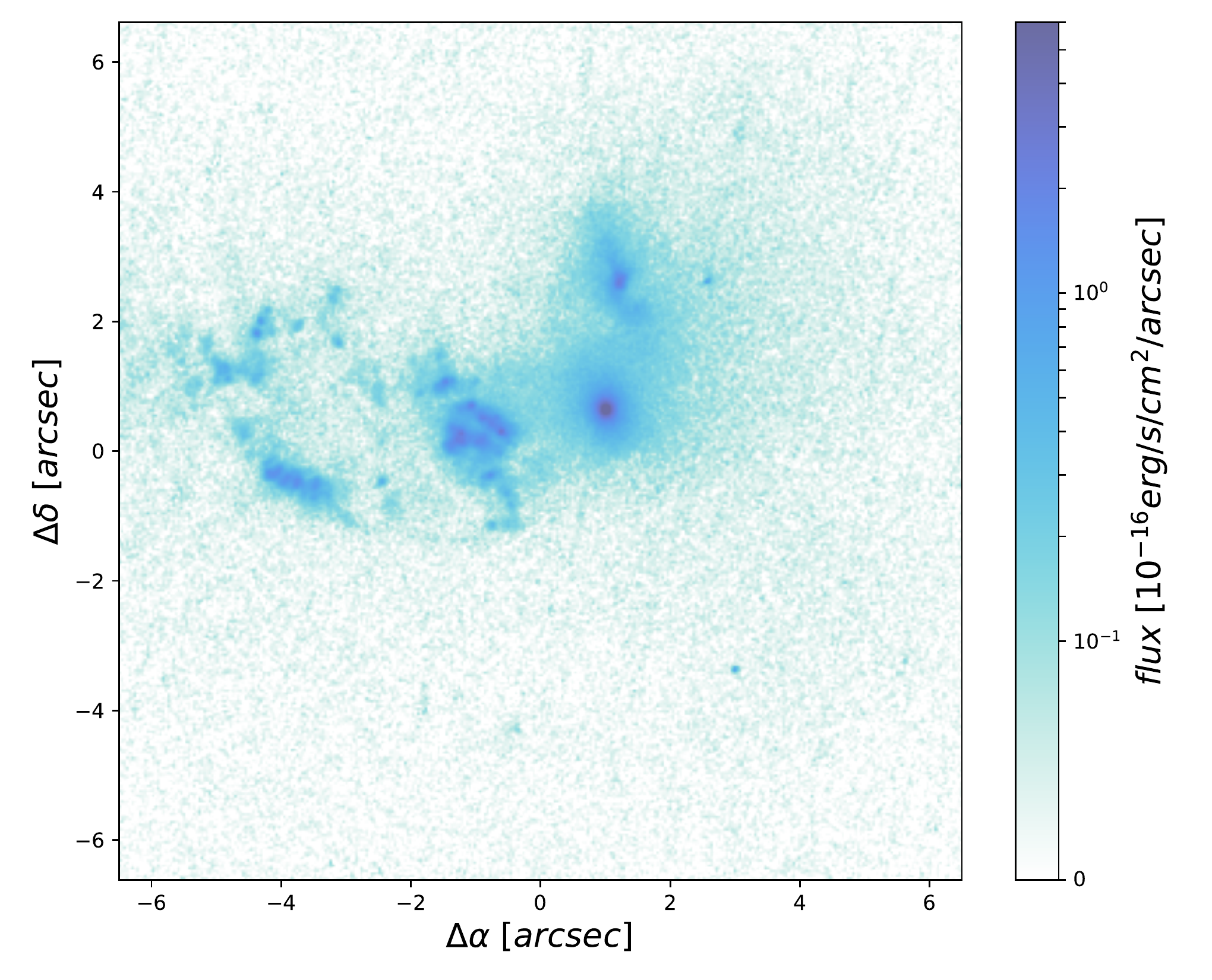}
\includegraphics[width=1.0\columnwidth]{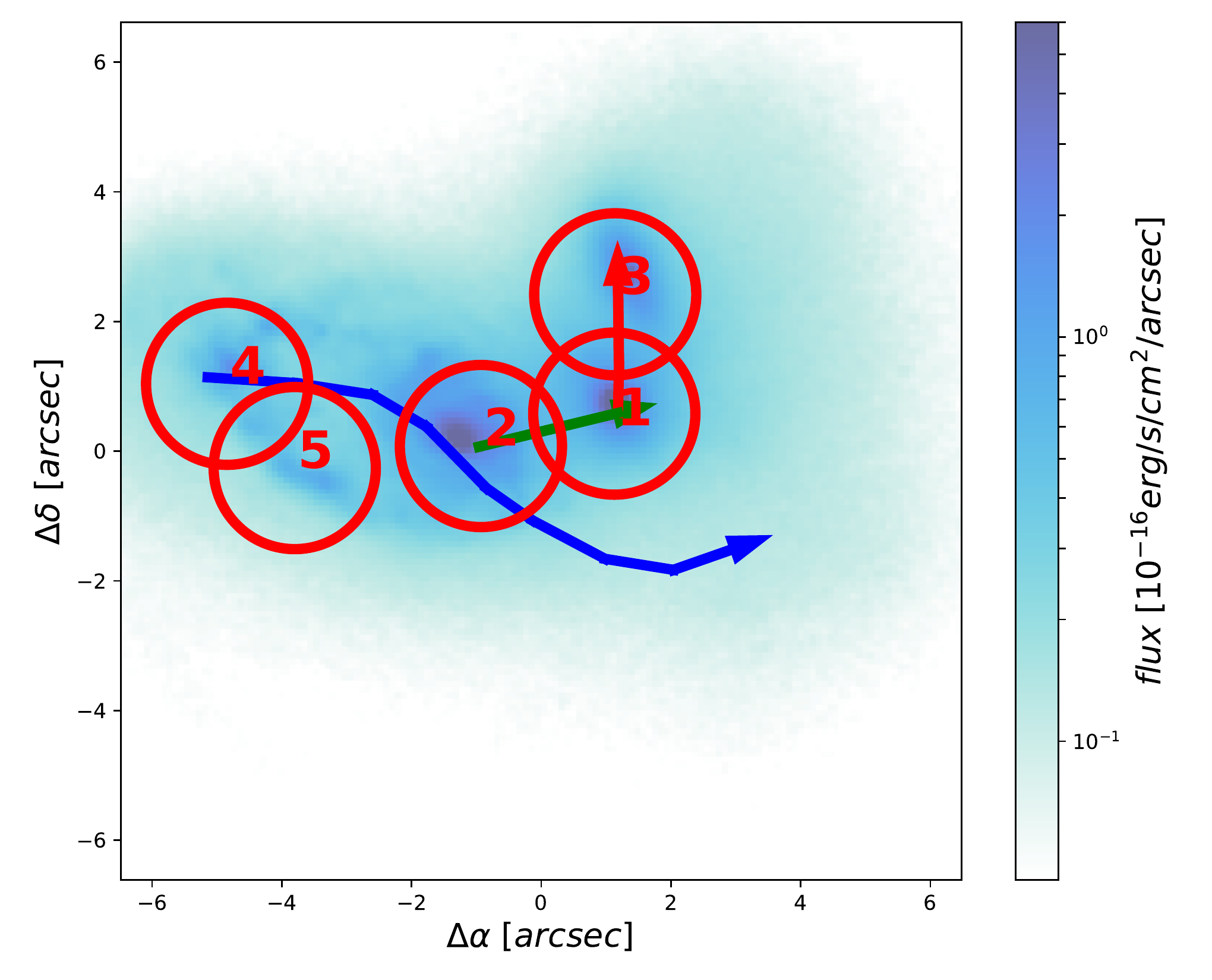}
\includegraphics[width=1.0\columnwidth]{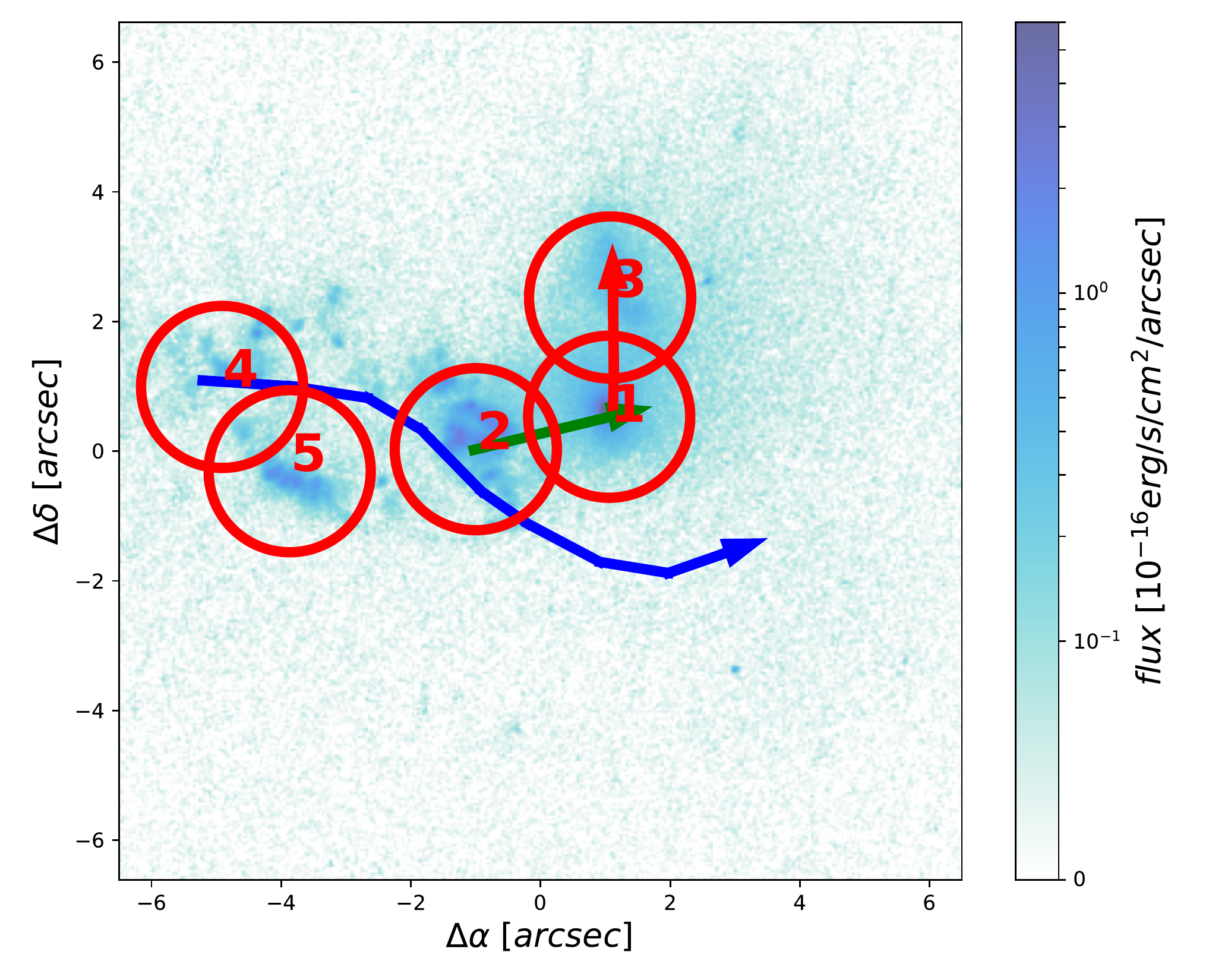}
\includegraphics[width=1.0\columnwidth]{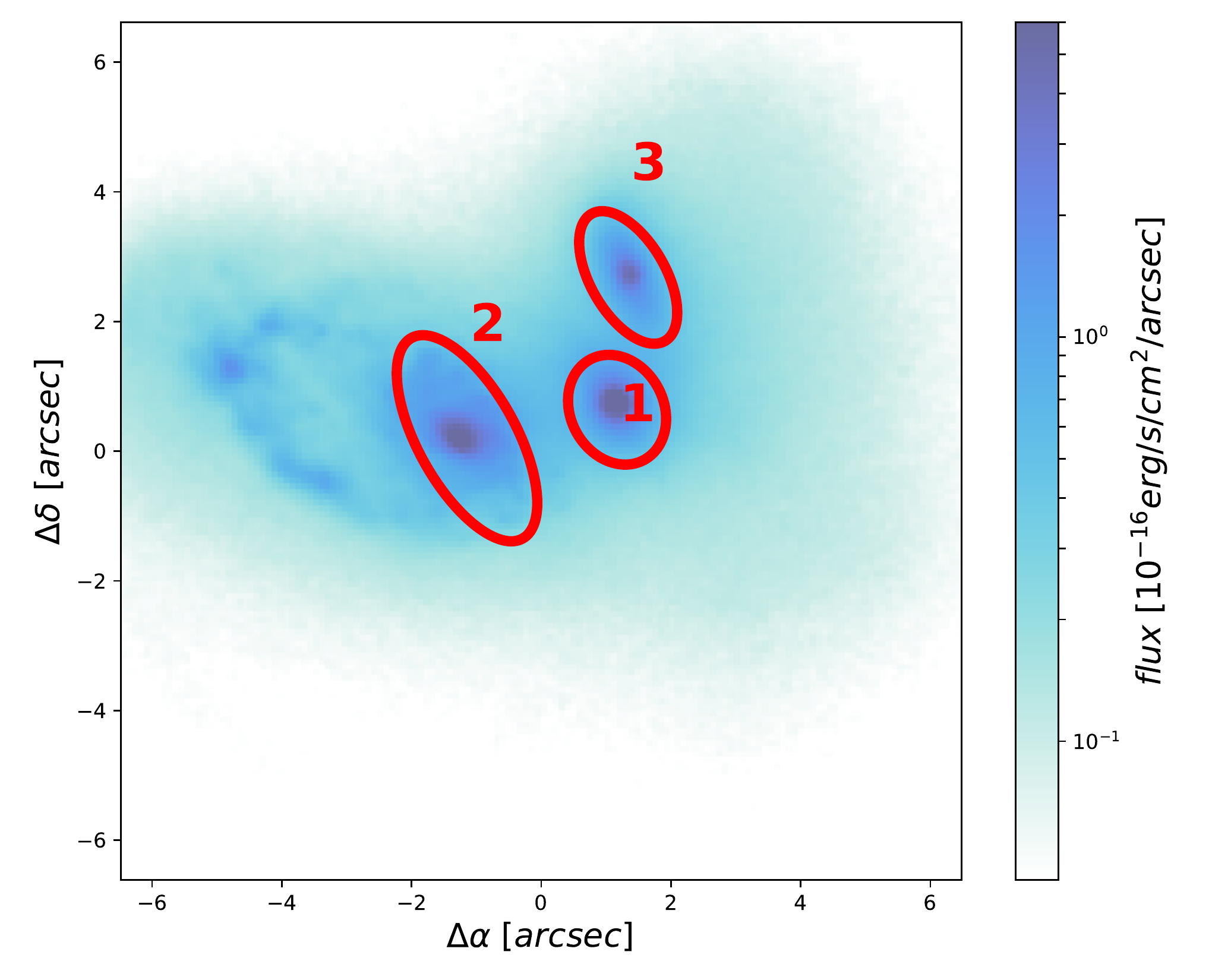}
\includegraphics[width=1.0\columnwidth]{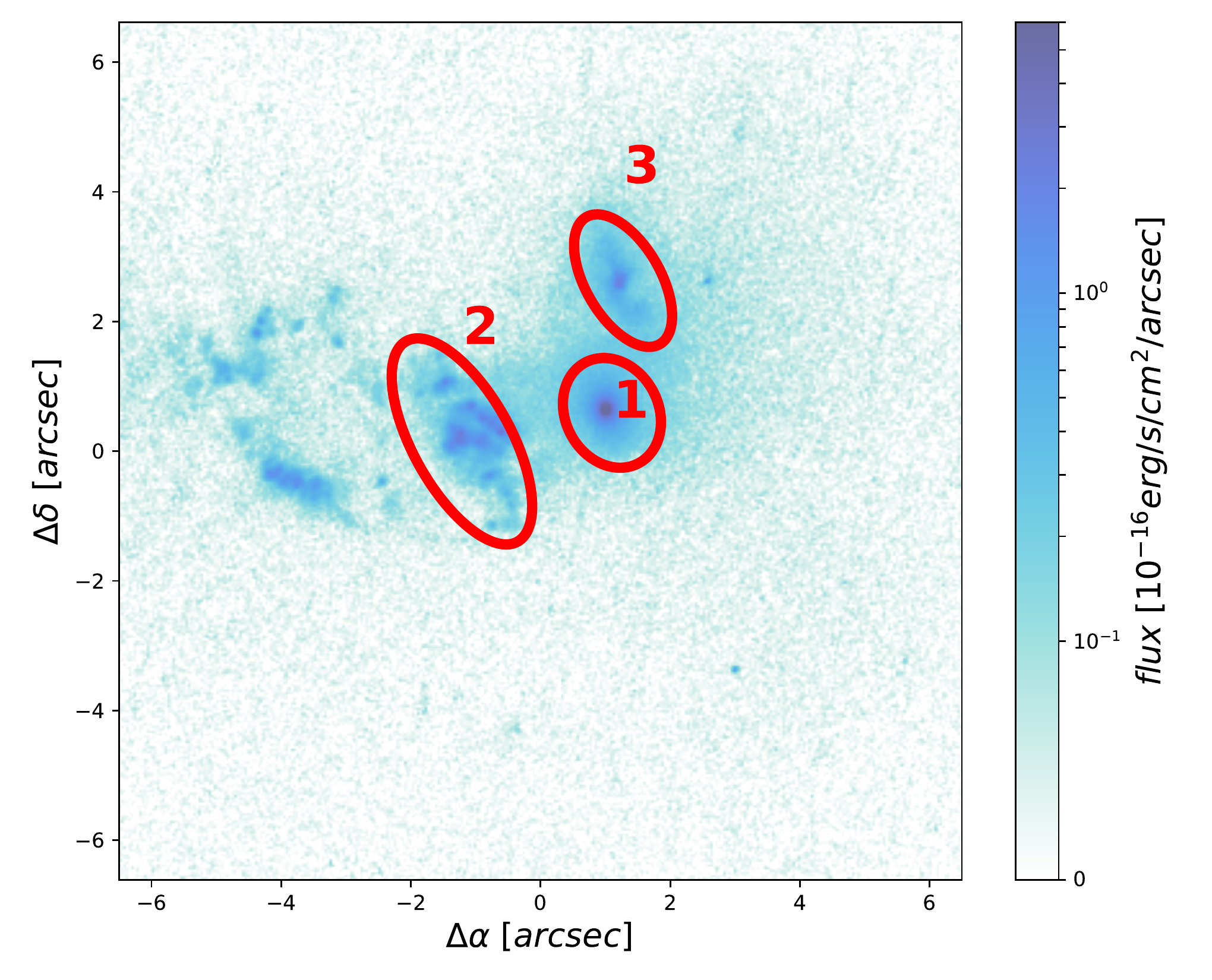}
\caption{
Left panels: The HST $Y$-band image of \J1027\ within the FoV of \meg. Right panels: HST $U$-band image of \J1027\ within the FoV of \meg. The upper panels show the HST images without any over-plotted regions. The middle panels show the HST images with the regions one to five, and also the plot of paths 1-3 (red arrow), 2-1 (green arrow), and A (blue arrow) of Figure \ref{havelmap}. The lower panels show the best-fit ellipse regions that contain at least the $50\%$ of the flux of regions 1, 2, and 3.}
\label{fig:hst}
\end{figure*}

The HST NIR image of \J1027 shows that regions 4 and 5, identified with \meg\ as SF regions, match in position with the knots found in the extended ring associated with \objedos, see Figure~\ref{fig:hst}. The HST optical-UV image of \J1027 also confirms that \objedos\ is disrupted, with a ring structure showing knots. 

These images allow us to fit the best ellipse region that encloses at least the $50\%$ of their total fluxes within their aperture region ($1.25\arcsec$ of radii). From these fittings, we measure that object \J1027\ (region one) has an elliptical shape with a semi-major axis of 0.89\arcsec (1.13\,kpc) and a semi-minor axis of 0.72\arcsec (0.92\,kpc). In the same way, object \objedos\ (region two) has a semi-major axis of 1.77\arcsec (2.26\,kpc) with a semi-minor axis of 0.76\arcsec (0.97\, kpc), and object \objetres\ (region three) has a semi-major axis of 1.13\arcsec (1.44\,kpc) with a semi-minor axis of 0.58\arcsec (0.74\,kpc).

\subsection{Dynamic mass}
\label{subsection:dynamass}

To calculate the dynamical mass for object \J1027, \objedos, and \objetres\ (regions one, two, and three), we apply the virial theorem \citep{Zwicky1937} with the implementation of the $S^2_K$ formulation \citep{Weiner+2006}, defined as:

\begin{displaymath}
M_{dyn}=\eta S^2_Kr_eG^{-1},
\end{displaymath}
with $S^2_K=\sigma^2+KV^2_{rot}$, $\sigma$ is the velocity dispersion, $V_{rot}$ the rotation velocity, and $r_e$ the effective radius of the object. In this case, we follow the analysis of \citet{Aquino+2018} and use the values of $K=0.5$, and $\eta=1.8$ to estimate the total dynamical mass for our three regions. To estimate the value of $r_e$, we use our measurements of the semi-major axis of the three objects. In addition, we set the value of the gravitational constant G as 4.3009$\times10^{-6}$\,kpc\,M$_{\odot}^{-1}$\,(\kms)$^2$. We also use the estimations of $\sigma_{\star}$ and $V_{max}$ to define $\sigma$ and $V_{rot}$ for the $S_K^{2}$ estimation. The final values of M$_{dyn}$, $\sigma_{\star}$ and $V_{max}$ are presented in Table \ref{table:main}.

\subsection{Black hole mass estimates} 
\label{bhmass}

We estimate the black hole mass of the three nuclei using the  M$_{BH}$ vs. $\sigma_{\star}$ correlation of \citet{2013ApJ...764..184M}:

\begin{displaymath}
    \text{log}_{10} \frac{M_{BH}}{M_{\odot}}=8.32+5.64\log_{10}\left(\frac{\sigma_{\star}}{200\,km s^{-1}} \right).
\end{displaymath}

We find that \J1027\ hosts a SMBH with a mass of $3.1\times10^9 M_{\odot}$, while \objedos\ have a SMBH mass of $1.9\times 10^9 M_{\odot}$ and \objetres\ have a SMBH mass of $1.7\times 10^9 M_{\odot}$.

\begin{deluxetable*}{lccccccccccc}
\tablenum{6}
\tablecaption{Main results}
\label{table:main}
\tablewidth{0pt}
\tablehead{
\colhead{Region} & \colhead{FWHM$_{B}$} & \colhead{FWHM$_{R}$} & \colhead{EW$_{B}$}& \colhead{EW$_{R}$} & \colhead{\mbh} & \colhead{M$_{\star}$} & \colhead{M$_{dyn}$} & \colhead{$\sigma_{\star}$} & \colhead{$V_{max}$} & \colhead{L$_X^{soft}$} & \colhead{L$_X^{hard}$}\\
\colhead{ } & \colhead{\kms} & \colhead{\kms} & \colhead{\AA}& \colhead{\AA} & \colhead{10$^{9}$M$_{\odot}$} & \colhead{10$^{10}$M$_{\odot}$} & \colhead{10$^{10}$M$_{\odot}$}& \colhead{\kms} & \colhead{\kms} & \colhead{10$^{40}$\ergs} & \colhead{10$^{40}$\ergs}
}
\decimalcolnumbers
\startdata
1 & 489$\pm$85 & 242$\pm$18 & 2.9$\pm$1.8  & 17.2$\pm$4.0 & 3.1$\pm$1.3 & 5.51$\pm$0.07 & 4.9$\pm$1.4  & 322$\pm$25 & $\ldots$ & 1.1$^{+0.8}_{-0.6}$ & 0.7$^{+1.3}_{-0.5}$\\
2 & 344$\pm$34 & $\ldots$   & 22.2$\pm$1.0  & $\ldots$  & 1.9$\pm$0.6 & 7.89$\pm$0.09 & 9.0$\pm$2.1   & 296$\pm$16 & 128$\pm$6 & 2.8$^{+0.9}_{-0.8}$ & 3.0$^{+1.6}_{-1.2}$\\
3 & 437$\pm$1  & 499$\pm$2  & 2.8$\pm$2.0  & 17.1$\pm$4.0 & 1.7$\pm$0.5 & 5.49$\pm$0.07 & 5.1$\pm$1.1  & 291$\pm$15 & $\ldots$ & 1.6$^{+0.8}_{-0.6}$ & 7.6$^{+5.9}_{-4.8}$\\
4 & 155$\pm$3  & $\ldots$   & 37.8$\pm$0.8  & $\ldots$  & $\ldots$    & 1.46$\pm$0.03 & $\ldots$ & 277$\pm$13 & $\ldots$ & $\ldots$ & $\ldots$\\
5 & 189$\pm$2  & $\ldots$   & 31.1$\pm$0.8  & $\ldots$  & $\ldots$    & 1.75$\pm$0.03 & $\ldots$ & 274$\pm$15 & $\ldots$ & $\ldots$ & $\ldots$\\
\enddata
\tablecomments{FWHM and EW(\ha) for the blue and red components only for \J1027\ and \objetres.}
\end{deluxetable*}

\subsection{WHAN diagram}
\label{whan}

\begin{figure*}[p]
  \sbox0{\begin{tabular}{@{}cc@{}}
  \includegraphics[width=1.2\textwidth]{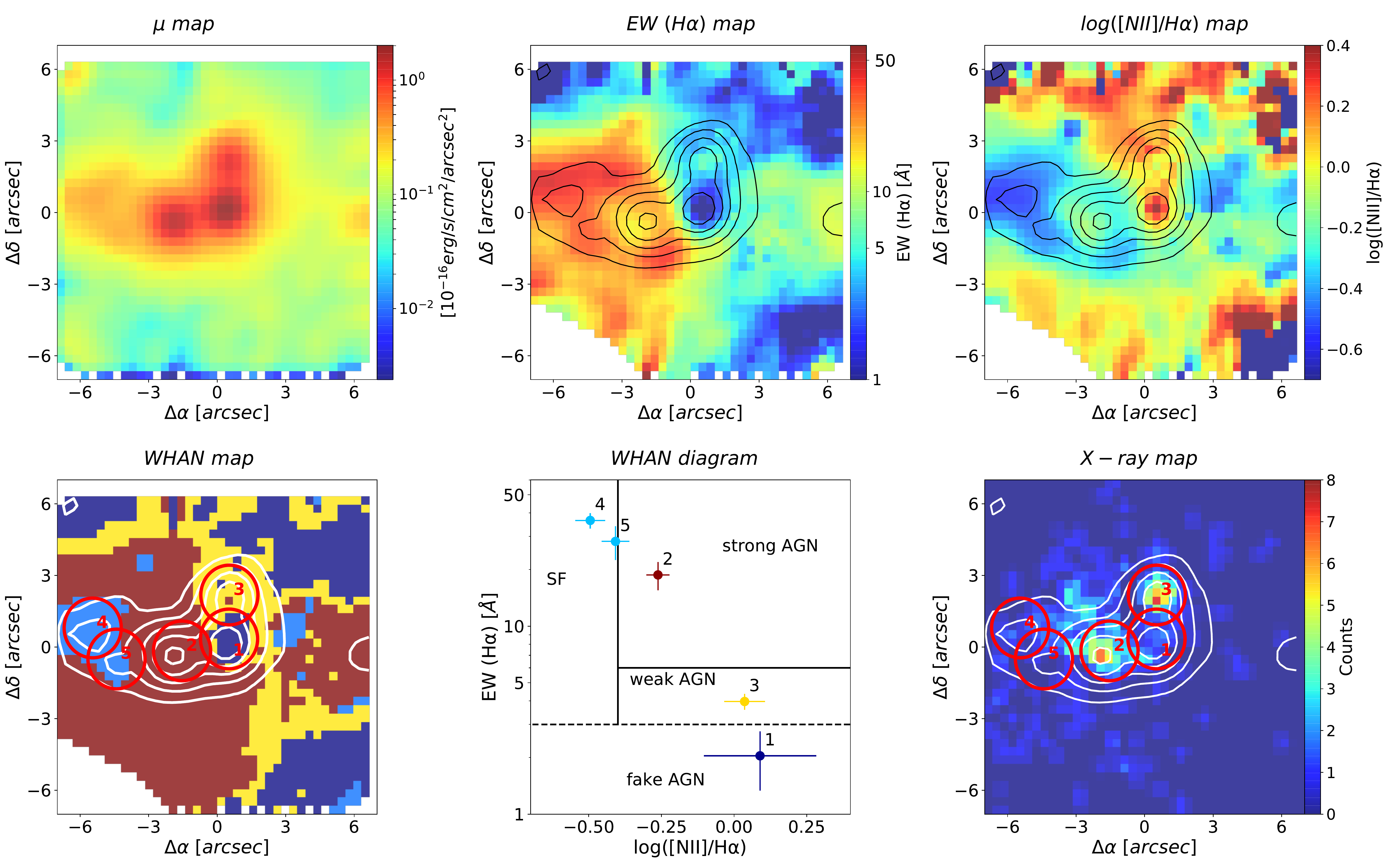}
  \end{tabular}}
  \rotatebox{90}{\begin{minipage}[c][\textwidth][c]{\wd0}
    \usebox0
    \caption{
Spatially resolved emission line analysis and X-ray map of \J1027. Upper left panel: Continuum surface brightness ($\mu$) map of the \meg\ FoV between the rest-frame wavelengths range of 6630 to 6670 \AA. Upper central panel: \ha\ equivalent width map within the \meg\ FoV, the solid black contours represent the continuum $\mu$ map for reference. Upper right panel: Resolved map of the logarithmic line ratio of \nii/\ha, the solid black contours represent the continuum $\mu$ map for reference. Lower left panel: Spatially resolved WHAN diagram within the \meg\ FoV. The light blue color zones represent the SF emission. The dark blue color represents the fake-AGN emission zones. The yellow color zones represent the weak-AGN emission. The red color zones represent the strong-AGN emission. The white contours represent the continuum $\mu$ map for reference and the red circles represent the five regions defined in this study. Lower central panel: The WHAN diagram obtained for the five regions, the color code is the same as shown in the Lower left panel. Lower right panel: X-ray emission within the \meg\ FoV from the Chandra 0.5-8 keV in counts. The red circles represent the five regions. The white contours represent the continuum $\mu$ map for reference. It is clear that the X-ray emission has very low counts in region 1, i.e. \J1027. \objedos\, shows extended X-ray emission towards the North with a size of $\sim$3.82 kpc.}
\label{fig:whan}
\end{minipage}}
\end{figure*}

With the 2D spatially resolved emission lines of Section \ref{2d_linefit}, we explore the resolved ionized emission properties along the \meg\ FoV. Hence, we obtain the map of \ha\ equivalent width EW(\ha). We use the continuum flux within 6630-6670\,AA\, to obtain the value of EW(\ha). We show the continuum map (upper left panel) and the EW(\ha) (upper middle panel) in Figure \ref{fig:whan}. In addition, in the upper right panel of Figure \ref{fig:whan}, we obtain the map of the line ratio  log\,\nii/\ha. In the EW(\ha) map, we find that in the locus of \J1027\ there is a depression of the EW(\ha) value with a minimum of $\approx$ 2 \AA, while \objedos\ has a larger value of 22 \AA, and \objetres\ an intermediate value of 4 \AA. Furthermore, the EW(\ha) distribution shows the existence of an extended structure centered in \objedos\ that matches with the structure discussed in Section \ref{section:HST_image} and the disk velocity profile discussed in Section \ref{subsection:rot}. Finally, in the locus of \J1027\, we find a peak on the log\,\nii /\ha\  map, with a maximum value of 0.2 dex.

We now proceed to explore the EW\ha\ vs. \nii /\ha\  diagram \citep[WHAN,][]{CidFernandes+2011}. On the lower left panel of Figure \ref{fig:whan}, we show the spatially resolved WHAN diagram. Following the definitions in \citet{CidFernandes+2011}, we separate the line rations into four regions with the next color code: The light blue zones represent the regions with emission within the star formation region in the WHAN diagram: log \nii/\ha\,$<$\,0.4 and EW(\ha)\,$>$\,3\,\AA. The dark blue zones represent the regions with emissions that characterize a fake AGN: EW(\ha)\,$<$\,3\,\AA. The yellow zones represent the regions within a weak AGN region in the WHAN diagram:  log \nii/\ha\,$>$\,-0.4 and 3 \AA $<$ EW(\ha)\ $<$ 6\AA. The dark red zones represent the regions with a strong AGN emission: log \nii/\ha\,$>$\,-0.4 and EW(\ha)\ $>$\,6\AA.

In addition, in the lower middle panel of Figure \ref{fig:whan}, 
we estimate the positions in the WHAN diagram of the integrated 
values of our five regions. Hence, the spatially resolved and 
integrated WHAN diagrams show that \J1027\ does not have any 
evidence of ionized gas emission that can be associated with an AGN or 
SF activity. The line ratios and EW(\ha) values indicate that \J1027\ 
can be classified as a fake AGN or as 
a ‘‘retired galaxy'' \citep[RG]{CidFernandes+2011}. 
The observed extended emission is associated with 
\objedos. This object shows a characteristic emission of a strong AGN emission 
\citep[Seyfert according with ][]{CidFernandes+2011}. Meanwhile, \objetres\ 
shows evidence of being a weak AGN in this diagram. For regions 4 and 5, 
we find the typical emission of SF regions. This conclusion is also 
corroborated in Section \ref{section:HST_image} identifying those regions 
as SF regions.

\subsection{The DPAGN in \J1027}

In Section \ref{sec:line_prof}, the detailed modeling of line emission profiles of the five apertures shows that only \J1027\ (region 1) and \objetres\ (region 3) have evidence of double-peaked narrow emission line components. This result would in principle suggests that both are DPAGN. 

On the other hand, from the spatially resolved kinematic analysis of Section \ref{sec:kinematic} and \ref{subsection:rot}, we show how the ionized extended 
emission from \objedos\ (region 2) partly propagates within the apertures of 
regions 1 and 3. We can summarize these effects as follows: a) Part of the 
ionized extended emission of \objetres\  and \objedos\ are collected within 
aperture 1 (\J1027). b) Part of the light from the ionized extended emission of 
\objedos\ is collected within aperture 3 (\objetres). c) The velocity shift from
the extended emission of \objetres\ is $\approx$ -150\,\kms at the boundaries 
of aperture 1. d) The velocity shift from the extended emission of \objedos\ 
is $\approx$ 40\,\kms at the boundaries of aperture 1, and is $\approx$ -150\,\kms 
at the boundaries of aperture 3.

\begin{deluxetable}{lcc}
\tablenum{7}
\tablecaption{Measured values of the blue and red components from the emission line profiles modeling within Regions 1 and 3. 
\label{table:fwhm-ew-lineratios}}
\tablewidth{0pt}
\tablehead{
\colhead{} & \colhead{Region 1} & \colhead{Region 3}  
}
\startdata
\ha$_{R}$ Flux     &    26$\pm$4    &   17$\pm$1    \\
\ha$_{B}$ Flux     &    17$\pm$4    &   27$\pm$2    \\
log \nii/\ha$_{R}$ & -0.16$\pm$0.15 & 0.03$\pm$0.15 \\
log \nii/\ha$_{B}$ &  0.01$\pm$0.10 & 0.11$\pm$0.10 \\
$<\Delta$V$_{R}>$  &    40$\pm$10   & -150$\pm$10   \\
$<\Delta$V$_{B}>$  &  -150$\pm$10   & -500$\pm$10   \\
\enddata
\tablecomments{Line fluxes in 10$^{-16}$erg\,s$^{-1}$\,cm$^{-2}$.
The blue and red  $<\Delta$V$>$ are in \kms.}
\end{deluxetable}

\begin{figure}
\centering
\includegraphics[width=0.9\columnwidth]{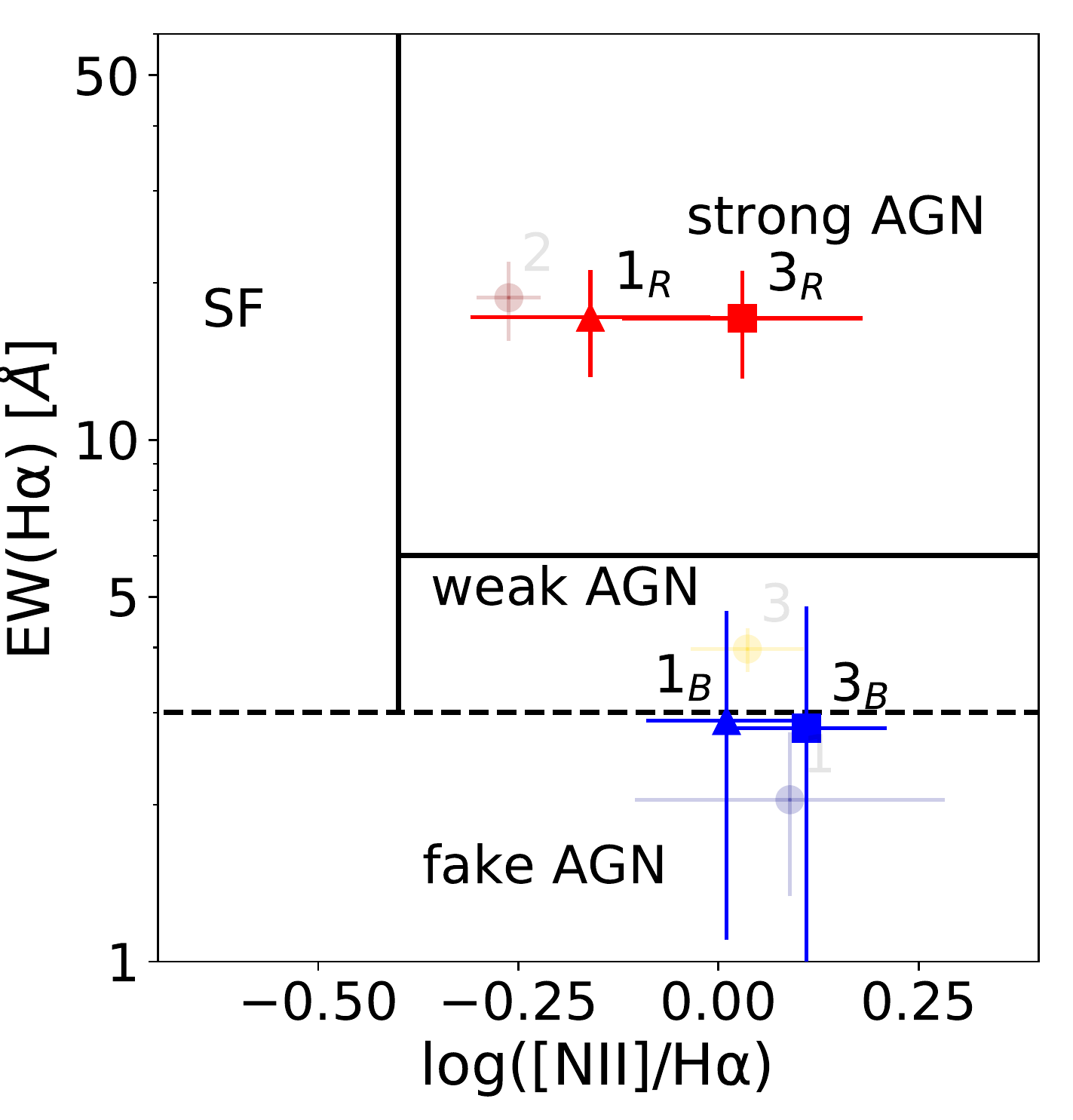}
\caption{WHAN diagram for the deblended blue and red components of the total spectra within regions 1 (triangles) and 3 (squares). The blue color represents the blue components, and the red color represents the red components. The positions of the averaged resolved values within apertures 1, 2, and 3 are overplotted with lighted colors following the color code of Figure~\ref{fig:whan}.}
\label{whan_blue_red}
\end{figure}

Therefore, the predicted velocity shifts from the spatially resolved kinematic maps are compatible with the velocity shifts of the blue and red components of the integrated spectra of \J1027\ and \objetres.
This support the hypothesis that the DPAGN observed in \J1027\ and \objetres\ are due to the extended emission  of \objedos\ into the apertures 1 and 3 (the red component), and a due to the extended emission of \objetres\ into aperture 1 (the blue component). 

To explore this point, we measure the position on the WHAN diagram of the decouple blue/red components of \J1027\ and \objetres, see Table \ref{table:fwhm-ew-lineratios}. We measure a log\nii/\ha\ line ratio for the \J1027\ red component of -0.16 $\pm$0.15, with a blue component of 0.01$\pm$0.1. For \objetres, we get a value of 0.03$\pm$0.15 for the red component and 0.11$\pm$0.1 for the blue component (see Table~\ref{table:fwhm-ew-lineratios}). However, to estimate the EW(\ha), we cannot directly measure the continuum flux for \J1027\ and \objetres\ because it is not possible to deblend the exact contribution of the continuum flux for each component. 

To turn around this problem, we assume that the light fraction of an emission line that is extended within a given aperture is the same as the light fraction of the continuum flux that extends to the same aperture. Therefore, we use the total flux of \ha\ of \objedos\ and the \ha\ flux of the red component of \objetres\ to re-scale the \objedos\ (aperture 2) continuum flux to estimate the continuum flux of the red component of \objetres\ (aperture 3): 
\begin{displaymath}
    Cont_{3,R}=Cont_{2}\frac{F(H\alpha_{3,R})}{F(H\alpha_{2})}.
\end{displaymath} The continuum flux associated to the blue component of \objetres\ is $Cont_{3,B}=Cont_{3}-Cont_{3,R}$. With this approximation, we estimate a continuum flux associated with the blue and red components within aperture 3 (\objetres). The red (blue) component of the continuum flux has a value of 1.0$\pm$2.0$\times10^{-16}$ \ergs cm$^{-2}$ (9.4 $\pm$2.0$\times10^{-16}$ \ergs cm$^{-2}$). Using the same approximation for aperture 1 (\J1027), we find a value of  1.5$\pm$2.0$\times10^{-16}$ \ergs cm$^{-2}$ for the red component, and a value of 5.9$\pm$2.0$\times10^{-16}$ for the blue component. With these values, we determine the value of EW(\ha) for each component. For \J1027\, we find an EW(\ha) value of 2.9$\pm$1.8 \AA\ for the blue component and a value of 17.2$\pm$4.0 \AA\ for the red component. For \objetres, we find a value of 2.8$\pm$2.0 \AA\ for the blue component and 17.1$\pm$4.0 \AA\ for the red component. We show all these values in Table \ref{table:main}.

We show the final position on the WHAN diagram in Figure \ref{whan_blue_red}. The positions of the red components of \J1027\ and \objetres\ are consistent with a strong AGN ionization, and are consistent with the ionization of \objedos. On the other hand, the blue component's positions on the WHAN are at the division between the weak AGN and fake AGN regions. However, once considering the errors, both components are consistent with the ionization of \objetres\ and \J1027. 

\subsection{X-ray data}

We have also analyzed archive data from Chandra Advanced CCD Imaging Spectrometer (ACIS-S) 0.5-8\,keV X-ray of \J1027. This observation was taken by the Chandra Program GO3-14104A (PI: Liu), with the ObsID=14971. These data were observed in 2013 01 23, and are part of an observational program that studies a set of triple AGN candidates \citep{Liu+19}.

We trim the Chandra X-ray image to fit the FoV of \meg\, and show the counts distribution map in the lower right panel of Figure \ref{fig:whan}. The X-ray compact emission is  astrometrically coincident with the optical positions of \objedos\ and \objetres. On the locus of \J1027, we observe that there are too few counts, so practically this nucleus does not show X-ray emission, supporting the results obtained with the WHAN diagram. Therefore, it is an object without evidence of AGN activity in these bands, in agreement with recent studies done by \citep[][]{Hou+2020,2021ApJ...907...71F,2021ApJ...907...72F}. On the other hand, \objedos\ and \objetres\ do show X-ray emission. Furthermore, \objedos\ has a clearly extended emission that coincides with part of the extended emission observed towards the NE in EW(\ha).

In order to estimate the X-ray fluxes, we have used the official Chandra Analysis tool {\sc CIAO}\footnote{Chandra Interactive Analysis of Observations: \url{https://cxc.cfa.harvard.edu/ciao/}} to obtain the calibrated soft (0.5-2 keV) and hard (2-8 keV) luminosities within the 2.5\arcsec apertures defined in this study.   

We use the tool {\sc srcflux}\footnote{\url{https://cxc.cfa.harvard.edu/ciao/ahelp/srcflux.html}} from the {\sc CIAO} package. With {\sc srcflux} we estimate the soft and hard X-ray net counts and fluxes from the Chandra X-ray observations. After net counts extraction, we use a power-law model \footnote{\url{https://cxc.cfa.harvard.edu/sherpa/ahelp/xspowerlaw.html}} to estimate the total flux. In addition, {\sc srcflux} uses a Bayesian computation to estimate the confidence intervals and errors from the extracted fluxes and counts. For the soft X-rays, we get 13$\pm$3 ($F_{soft}$=1.1$^{+0.7}_{-0.6}$), 31$\pm$6  ($F_{soft}$=2.6$^{+0.9}_{-0.7}$), and 17$\pm$4  ($F_{soft}$=1.5$^{+0.7}_{-0.6}$) counts (fluxes 10$^{-15}$ erg\,s$^{-1}$\,cm$^{2}$) for regions 1, 2 and 3 respectively. In the same way, we obtain 2$\pm$1 ($F_{hard}$=0.6$^{+1.2}_{-0.5}$), 9$\pm$3 ($F_{hard}$=2.8$^{+1.5}_{-1.1}$), and 23$\pm$5  ($F_{hard}$=7.1$^{+5.6}_{-4.5}$) counts (fluxes 10$^{-15}$\,erg\,s$^{-1}$\,cm$^{2}$) for the hard X-rays for Regions 1, 2 and 3. The background contribution is almost negligible with $\approx$0.08 counts for the soft X-rays and $\approx$0.2 counts for the hard X-rays. Finally, the soft and hard X-ray luminosities are shown in Table~\ref{table:main } by using a luminosity distance with $z=0.066$.
\section{Discussion}
\label{section:dis}

Multiple AGN systems are expected to be found in the framework of the hierarchical model of galaxy formation. Since the source \J1027\ is among the few triple AGN candidate systems known up to date, GTC/\meg\ observations result ideal to study the nature and kinematics at kpc scales of these nuclei that are associated within this field. The spatially resolved data cube image obtained with \meg\ allows the establishment of the extension of the ionized emission gas and the kinematical properties of each nucleus. The WHAN diagram was chosen to provide the classification of the nuclei since the blue part of the \meg\ spectra could not be used in this analysis. This is due to the high obscuration and low SNR obtained in this part of the spectrum, so the BPT diagnostic diagrams could not be obtained. 

Our stellar mass estimation of \J1027 was compared with the ones obtained by L11 and \citet[C15][]{2015ApJS..219....8C}. In our case, we have considered the stellar mass correction from a Chabrier/Kroupa IMF to a Salpeter IMF\footnote{We use a scale factor of 0.21 dex and 0.20 dex to go from a Chabrier/Kroupa to a Salpeter IMF \citep{Madau+14}}. So, we find a stellar mass value of $5.51\times10^{10}M_{\odot}$. On the other hand, L11 obtained a stellar mass of $2.6\times10^{11}M_{\odot}$, while C15 a value of $1.7\times10^{11}M_{\odot}$. The smaller value obtained by us could be due to differences in the total aperture size, and in the way that {\sc pyPipe3D} calculates the internal dust extinction. For instance, if we calculate the total stellar mass within the SDSS fiber aperture, we find a total value of 8.6$\pm$0.1 $\times10^{10}M_{\odot}$ for \J1027. In addition, the stellar masses estimated from the Spectral Energy Distribution (SED) fitting, whereas it uses the optical spectra (the case of L11) or a panchromatic SED fitting (C15) are subject to large systematic uncertainties due to a complex mix of dust extinction, stellar libraries, inclination, etc \citep[see ][]{Ibarra-Medel+2019}.

The WHAN diagram shows that \J1027, previously identified as a LINER by L11, is instead a retired galaxy that does not have activity produced by either AGN or SF. Also, we find that \J1027\ is a DPAGN based on our analysis of the 1D SDSS spectrum shown in Figure~\ref{fig:megmodelA}. However, \meg\ data analysis clearly shows that the origin of the double-peaked narrow emission lines in \J1027\ is the result of the extended narrow emission coming from \objedos, and also of weak AGN emission from \objetres\ (see upper right panel of Figure~\ref{havelmap}). Therefore, we have a case where a DPAGN classification leads us to misclassify an RG as a DPAGN. Meanwhile, the DPAGN observed within region 3 (\objetres) is due to the ionized extended emission of \objedos, plus its intrinsic ionized emission, so in this case, \objetres\ is a DPAGN. 

The spatially resolved WHAN diagram shows that the source of the ionization emission in \J1027\ is produced by Hot Low-mass evolved stars \citep[HOLMES][]{Stasinska+04,CidFernandes+2011}. This is also supported by the high value of log\,\nii/\ha\ (0.2 dex) and the low value of EW(\ha) found in the WHAN diagram for \J1027. Therefore, it is easy to misclassify an RG as a LINER-like object using only the BPT diagrams and as a DPAGN using long-slit spectroscopy. 

This result is also confirmed by revisiting the Chandra X-ray data of \J1027. The X-ray counts map shows basically no emission being produced by this RG. The HST optical-UV and NIR images show it has an elliptical morphology, typical of RGs. So, our analysis suggests that \J1027\ is not an AGN. Therefore, these results rule out the triple AGN nature of the triplet of galaxies, at least in the optical and X-ray bands.

In the WHAN diagram, \objedos\ appears as a strong AGN. Due to its redshift, we classify it as a Seyfert 2 (Sy2) galaxy. The Sy2 classification is also supported by the results obtained from the MCMC modeling of the spaxels of the data cube. The model shows that no broad emission line components appear to be present in either of the three nuclei within 1-$\sigma$ confidence level. The iso velocity map shows that \objedos\ has a rotating gaseous disk, and the WHAN diagram confirms the presence of kpc-scale extended \ha\ emission region associated with this Sy2. This galaxy was previously classified as a composite-AGN by L11. 

The X-ray data for \objedos\ show emission at 0.5-8 keV, and also extended emission with a size of $\sim$3.8\,kpc that coincides with the NE extended emission observed in the WHAN diagram. HST images reveal that \objedos\ shows signs of interaction, with a disrupted ring that shows emission from SF regions located at regions 4 and 5, shown in the WHAN diagrams. We find that the l.o.s. velocity difference between \objedos\ and \J1027 is only $\approx$ 4\,\kms. 

In the case of \objetres, it was found to be a weak AGN, in this case, a LINER2 \citep[e.g.][]{2000ApJ...533..729T}. The HST NIR and optical-UV images show that this object has a spiral morphology, although the HST images show it with wider opened spiral arms produced by the interaction with \objedos\ and \J1027. In fact, with our data, we find that \objetres\ is moving with respect to \objedos\ and \J1027\ with a velocity of $\approx$-500\,\kms. 

\section{Conclussions}
\label{section:conc}

The kinematical analysis presented in this work shows that the Northern nucleus is moving with respect to the South-Eastern nucleus with a speed of $\sim$-500\,\kms. This large velocity suggests that both AGN are in a late minor merger phase. We have also found that both are obscured objects, in agreement with L11 results. If the Northern nucleus passes near the center of the disk of the South-Eastern nucleus, after n passings, it can produce a collisional ring with knots, a morphology that is observed in the NIR, optical-UV HST images, and also in agreement with results from numerical simulations \citep[e.g.][]{1977ApJ...212..616T,2018MNRAS.473..585R}. Since the Northern nucleus is the less massive companion galaxy, and interacts with the South-Eastern nucleus with a head-on collision, the latter object, is predicted to end up with lenticular morphology as a consequence of this kind of interaction \citep[][]{,1998ApJ...499..635B,2023MNRAS.tmp..480G}. 

Finally, we have shown that the former triple AGN candidate in \J1027\ is instead a DAGN system where the central nucleus of the system, is not an AGN. The three nuclei are characterized by the spatially resolved WHAN diagram, the center nucleus a Retired Galaxy, the Northern nucleus a LINER2, and the South-Eastern nucleus a Sy2 galaxy. So, the DAGN system consists of a pair of Sy2-LINER2, with a projected separation of 3.98\,kpc. This DAGN system is confirmed in the optical and X-ray data analyzed in this work.

\begin{acknowledgments}
We thank the anonymous referee for their valuable comments and suggestions. EB,ICG and JMRE acknowledge support from DGAPA-UNAM grant IN113320. ICG,EB,HIB,CAN,JMRE acknowledge support from DGAPA-UNAM grant IN119123. CAN thanks support from DGAPA-UNAM grant IN111422 and CONACyT project Paradigmas y Controversias de la Ciencia 2022-320020. HIM acknowledges the  support from the Instituto Polit\'ecnico Nacional (M\'exico). JMRE acknowledges the Spanish State Research Agency under grant number AYA2017-84061-P. JMRE also acknowledges the Canarian Government under the project PROID2021010077,  and is indebted to the Severo Ochoa Programme at the IAC. X.L. acknowledges support by NSF grant AST-2108162. Based on observations made with the Gran Telescopio Canarias (GTC), installed at the Spanish Observatorio del Roque de los Muchachos of the Instituto de Astrofísica de Canarias, on the island of La Palma. This work is based on data obtained with \meg\ instrument, funded by European Regional Development Funds (ERDF), through Programa Operativo Canarias FEDER 2014-2020.   We acknowledge the support given by GTC-{\it MEGARA} staff, in particular to Antonio Ca\-bre\-ra. This work is partly based on observations made with the NASA/ESA Hubble Space Telescope, obtained from the data archive at the Space Telescope Science Institute. STScI is operated by the Association of Universities for Research in Astronomy, Inc. under NASA contract NAS 5-26555. Funding for the Sloan Digital Sky Survey V has been provided by the Alfred P. Sloan Foundation, the Heising-Simons Foundation, the National Science Foundation, and the Participating Institutions. SDSS acknowledges support and resources from the Center for High-Performance Computing at the University of Utah. The SDSS website is www.sdss.org. Support for Chandra program number 14700279 was provided by NASA through Chandra Award Number GO3-14104X issued by the Chandra X-ray Observatory Center, which is operated by the Smithsonian Astrophysical Observatory for and on behalf of NASA under contract NAS 8-03060. This research has made use of the NASA/IPAC Extragalactic Database (NED), which is operated by the Jet Propulsion Laboratory, California Institute of Technology, under contract with the National Aeronautics and Space Administration. 

\end{acknowledgments}

\clearpage
\bibliography{bibliography}
\addcontentsline{toc}{chapter}{Bibliography}

\facilities{GTC/(\meg\-IFU), CXO (ACSIS), HST (WFC3), Sloan Digital Sky Survey Legacy DR7.}


\software{\meg\ DRP \url:https://megaradrp.readthedocs.io/en/stable/,
          PYPIPE3D \citep{Sanchez+2022},
          IRAF/specfit \citep{1994ASPC...61..437K}. 
         }

\end{document}